\documentclass[11pt,osid]{article}
\usepackage{amsmath}
\usepackage{graphicx}
\usepackage{amsfonts}
\usepackage{subfigure}
\usepackage{bbm,footmisc}

\newtheorem{theorem}{Theorem}[section]

\DeclareMathAlphabet\mathbfcal{OMS}{cmsy}{b}{n}

\newcommand{\R}{\mathbb{R}}

\newcommand{\rr}{\mathbb{R}}

\newcommand{\sy}[1]{{\rm Sp}{(#1,\R)}}
\newcommand{\id}{\mathbb{I}}
\renewcommand{\text}[1]{{\rm #1}}
\newcommand{\hh}{\mathcal{H}}
\newcommand{\T}{^{\sf T}}
\newcommand{\gr}[1]{\boldsymbol{#1}}

\newcommand{\be}{\begin{equation}}
\newcommand{\ee}{\end{equation}}
\newcommand{\bea}{\begin{eqnarray}}
\newcommand{\eea}{\end{eqnarray}}
\newcommand{\eq}[1]{Eq.~(\ref{#1})}

\newcommand{\tr}{{\rm Tr}\,}
\newcommand{\ket}[1]{|#1\rangle}
\newcommand{\bra}[1]{\langle#1|}
\newcommand{\ketbra}[2]{\vert #1 \rangle \! \langle #2 \vert}

\newcommand{\op}[1]{\hat{#1}}
\newcommand{\adj}[1]{{#1}^{\dag}}
\newcommand{\comm}[2]{\left[#1,#2\right]}
\newcommand{\sig}{{\gr\sigma}}
\newcommand{\eps}{\gr{\varepsilon}}

\newcommand{\gam}{\boldsymbol{\gamma}}

\begin{document}
\title{Continuous variable quantum information: Gaussian states and beyond}

\author{Gerardo Adesso, Sammy Ragy, and Antony R. Lee \\{\footnotesize\it School of Mathematical Sciences, The University of Nottingham}, \\ {\footnotesize\it University Park, Nottingham NG7 2GD, United Kingdom} \\ {\footnotesize gerardo.adesso@nottingham.ac.uk}}

\maketitle

\begin{abstract}
The study of Gaussian states has arisen to a privileged position in continuous variable quantum information in recent years. This is due to vehemently pursued experimental realisations and a magnificently elegant mathematical framework. In this article, we provide a brief, and hopefully didactic, exposition of Gaussian state quantum information and its contemporary uses, including sometimes omitted crucial details. After introducing the subject material and outlining the essential toolbox of continuous variable systems, we define the basic notions needed to understand Gaussian states and Gaussian operations. In particular, emphasis is placed on the mathematical structure combining notions of algebra and symplectic geometry fundamental to a complete understanding of Gaussian informatics. Furthermore, we discuss the quantification of different forms of correlations (including entanglement and quantum discord) for  Gaussian states, paying special attention to recently developed measures. The manuscript is concluded by succinctly expressing the main Gaussian state limitations and outlining a selection of possible future lines for quantum information processing with continuous variable systems.
\end{abstract}

\section{Introduction}\label{secIntro}
Quantum information technology has reached remarkable milestones in the last three decades \cite{BNielsChuang} and promises even more revolutionary advances in the next three \cite{qinternet}. Pioneering proposals of the likes of quantum cryptography \cite{BB84,Ekert91} and teleportation \cite{Telep} have now been demonstrated in countless experiments with a variety of quantum hardware, and entered a stage of commercial exploitation \cite{idquantique}. The time seems ripe for selected quantum computing devices to live up to their high expectations \cite{dwave}.

Traditionally, two main approaches to quantum information processing have been pursued. On one hand, a ``digital'' one, according to which information is encoded in systems with a discrete, finite number of degrees of freedom, so-called qubits or qudits. Typical examples of qubit implementations are the nuclear spins of individual atoms in a molecule, the polarisation of photons, ground/excited states of trapped ions, etc. In parallel, an ``analog'' approach has also been devised, based on quantum information and correlations being encoded in degrees of freedom with a continuous spectrum ({\it continuous variables}), such as those associated to position and momentum of a particle. This second approach has witnessed considerable success due to its versatility, with implementations often encompassing different physical systems, e.g.~light quadratures and collective magnetic moments of atomic ensembles, which obey the same canonical algebra.

The aim of this short article is to offer a tour of the recent progress in continuous variable (CV) quantum information, and to get an interested reader started on the mathematical formalism suitable for the characterisation of the subject. There are already quite a good number of, more or less up-to-date, bibliographic resources on quantum information with CVs; see e.g.~\cite{sammyrev,eisplenio,BBraunstein,parisbook,brareview,book,ourreview,ciracreview,pirandolareview}. This contribution absolutely does not strive to provide a comprehensive coverage of the field, but is more oriented towards a didactic exposition of some basic concepts which may be overlooked in denser review articles, and a teaser of future perspectives and currently open problems.

Particular emphasis will be given on a special subfield of CV quantum information, which revolves around the use of {\it Gaussian states} and {\it operations}.
Gaussian states constitute versatile resources for quantum communication protocols with bosonic CV systems \cite{sammyrev,parisbook,brareview,book,pirandolareview} and are important testbeds for investigating the structure of quantum correlations \cite{ourreview}, whose role is crucial in fields as diverse as quantum field theory, solid state physics, quantum stochastic processes, and open system dynamics. Gaussian states naturally occur as ground or thermal equilibrium states of any physical quantum system in the `small-oscillations' limit \cite{ciracreview,extra}. Moreover, some transformations, such as those associated with beam splitters and phase shifters, as well as noisy evolutions leading to loss or amplification of quantum states, are naturally Gaussian, that is, they map Gaussian states into Gaussian states. Gaussian states are furthermore particularly easy to prepare and control in a range of setups including primarily quantum optics, atomic ensembles, trapped ions, optomechanics, as well as networks interfacing these diverse technologies  \cite{book}. From the mathematical perspective, Gaussian states are technically accessible, since they are completely described by a finite number of degrees of freedom only (first and second moments of the canonical mode operators), despite their infinite-dimensional support. Their description in a quantum phase space picture, thanks to the symplectic formalism  \cite{Arvind95}, is particularly effective, as we will demonstrate in the following.

The paper is organised as follows.
In Section~\ref{secCV} we introduce CV systems and their phase space description. In Section~\ref{secGauss} we focus on Gaussian states and operations and characterise their informational properties. In Section~\ref{secGaussEnt} we give an overview on entanglement and other measures of correlations for Gaussian states of multimode CV systems.  In Section~\ref{secBeyond} we discuss some limitations of Gaussian resources for CV quantum protocols and wrap up with a summary and outlook.

In this article we adopt a number of notation conventions. Operators on a Hilbert space will be denoted with a hat, $\hat{\cdot}$, physical quantum states as $\rho$, the transposition and Hermitian conjugation operations as $\cdot\T$ and $\cdot^{\dag}$ respectively. Complex conjugate by an overline, $\overline{\,\,\cdot\,\,}$, the determinant of a matrix as $\det(\cdot)$ and the tracing operation $\mathrm{Tr}(\cdot)$. Further, we will denote both vectors and matrices with boldface, e.g.~$\boldsymbol{\xi}$ and $\boldsymbol{\sigma}$; $\gr{I}$ will denote the identity matrix. A matrix is positive semidefinite, $\gr{M} \geq 0$, when all of its eigenvalues are nonnegative. Finally, we will be adopting the ubiquitous natural units convention $\hbar=c=1$; notice that other  references may assume different conventions \cite{pirandolareview}.

\section{Continuous variable systems}\label{secCV}

\subsection{Quantised fields, modes, and canonical operators}
We can define a CV system, in quantum mechanics, as a system whose relevant degrees of freedom are associated to operators with a continuous spectrum. The eigenstates of such operators form bases for the infinite-dimensional Hilbert space ${\cal H}$ of the system. An archetypical bosonic CV system is a quantised field, such as the electromagnetic field. Such a system can be modeled as a collection of non-interacting quantum harmonic oscillators with different frequencies; each oscillator is referred to as a {\it mode} of the system, and is a CV system in its own right. To avoid complications associated with quantum field theory, we will consider systems with a discrete number of modes (e.g., an optical cavity), each with one spatial dimension.
Mathematically, then, a CV system of $N$ canonical bosonic modes
is described by a Hilbert space $\hh=\bigotimes_{k=1}^{N} \hh_{k}$
resulting from the tensor product structure of infinite-dimensional
Hilbert spaces $\hh_{k}$'s, each of them associated to a single mode \cite{eisplenio,brareview,ourreview}.
In the case of the electromagnetic field, the Hamiltonian of the complete system can be written as
\begin{equation}\label{CV:Ham}
\op{H} = \sum_{k=1}^N \op{H}_k\,,\quad \op{H}_k=\hbar \omega_k \left(\adj{\op{a}}_k\op{a}_k +
\frac12\right)\,,
\end{equation}
where each term $\op{H}_k$ in the sum refers to the reduced Hamiltonian of the $k^{\rm th}$  mode of the field (a single harmonic oscillator).
Here $\op{a}_k$ and $\adj{\op{a}}_k$ are the  annihilation and
creation operators of an excitation in mode $k$ (with frequency
$\omega_k$), which satisfy the canonical commutation relation
\begin{equation}\label{CV:comm}
\comm{\op{a}_k}{\adj{\op{a}}_{l}}=\delta_{kl}\,,\quad
\comm{\op{a}_k}{\op{a}_{l}}=\comm{\adj{\op{a}}_k}{\adj{\op{a}}_{l}}=0\,.
\end{equation}
Adopting natural units, the corresponding quadrature phase operators (`position'
and `momentum')  for each mode are defined as
\be\label{qpqp}  \hat q_{k} = \frac{(\op a_{k}+\op a^{\dag}_{k})}{\sqrt{2}}\,, \quad
  \hat p_{k} = \frac{(\op a_{k}-\op a^{\dag}_{k})}{i \sqrt 2}\,.
\ee
and satisfy the bosonic commutation relations $[\op{q}_k,\op{p}_l]=i \delta_{kl}$.
We can group together the canonical operators in the vector
\be\label{CV:R}
{\hat {\gr R}}=(\hat{q}_1,\hat{p}_1,\ldots,\hat{q}_N,\hat{p}_N)\T\,,
\ee
which enables us to write in compact form the  bosonic commutation
relations between the quadrature phase operators, \be
[\hat{R}_k,\hat{R}_l]= i\,\Omega_{kl} \; ,\label{ccr}\ee where
$\boldsymbol{\Omega}$ is the $N$-mode symplectic form \be
\boldsymbol{\Omega}=\bigoplus_{k=1}^{N}\boldsymbol{\omega}\, , \quad  \boldsymbol{\omega}=
\left(\begin{array}{cc}
0&1\\
-1&0
\end{array}\right)\, . \label{eqn:real-symplectic-form}
\ee

The space $\hh_k$ for each mode $k$ can be spanned by the Fock basis $\{\ket{n}_k\}$ of
eigenstates of the number operator $\hat{n}_k = \hat a_k^{\dag}\hat a_k$, which are also the eigenstates of the Hamiltonian $\op{H}_k$ of the noninteracting mode, see \eq{CV:Ham}. The Hamiltonian of each mode is bounded from below, thus
ensuring the stability of the system. For each mode  $k$
there exists a  vacuum state $\ket{0}_k\in \hh_k$
such that \be \label{vuoto} \hat a_k\ket{0}_k=0.\ee
The vacuum state of the complete $N$-mode system will be denoted by
$\ket{0}=\bigotimes_k \ket{0}_k \in {\cal H}$.
Other bases for ${\cal H}_k$ can be the continuous ones corresponding to eigenstates of the position or momentum operators, i.e.~plane waves (although these are quite unpractical). Another useful alternative basis is provided by {\it coherent states} \cite{BWallsMilburn}, which are the right-eigenstates of the annihilation operator $\hat a_k$ and form an overcomplete basis in $\hh_k$.
Coherent states $\ket{\alpha}_k$ result from applying the
single-mode Weyl displacement operator $\hat D_k$ to the vacuum
$\ket{0}_k$, $\ket{\alpha}_k= \hat D_k(\alpha)\ket{0}_k$, where
\be\label{CV:Weyl}\hat D_k(\alpha)=\,e^{\alpha \hat
a_k^{\dag}-\overline{\alpha}\hat a_k}\,,\ee and the coherent amplitude
$\alpha\in{\mathbb C}$ satisfies $\hat
a_k\ket{\alpha}_k=\alpha\ket{\alpha}_k$. In terms of the Fock basis
of mode $k$ a coherent state reads
\be\label{CV:coh}\ket{\alpha}_k=\,e^{-\frac12|\alpha|^2}
\sum_{n=1}^{\infty}\frac{\alpha^{n}}{\sqrt{n!}}\ket{n}_k\,.\ee
Tensor products of coherent states for $N$ different modes are obtained
by applying the $N$-mode Weyl operators $\hat D({\gr\xi})$ to the global
vacuum $\ket{0}$ defined above. For future convenience, we define the operators
$\hat D({\gr\xi})$  in terms of the canonical operators $\hat{\gr R}$, \be \label{weyn} \hat
D({\gr\xi}) = \,e^{i\hat{\gr R}^{\sf T} \boldsymbol{\Omega} \gr\xi}\, , \quad {\rm
with} \quad\gr\xi\in {\mathbb R}^{2N} \; . \ee
Subsequently, one has
$\ket{\gr\xi}=\hat D_{\gr\xi}\ket{0}$.
For a single mode $k$, notice that the definition (\ref{weyn}), $\hat{D}_k{{\xi_1 \choose \xi_2}} = e^{i(\xi_2 \hat{q}_k - \xi_1 \hat{p}_k)}$,  reproduces Eq.~(\ref{CV:Weyl}) upon setting $\boldsymbol{\xi}\equiv{{\xi_1 \choose \xi_2}} = \sqrt{2}{{\mathrm{Re}(\alpha)}\choose{\mathrm{Im}(\alpha)}}$ and exploiting Eqs.~(\ref{qpqp}).

\subsection{Phase space description}

The states of a CV system are the set of positive semidefinite trace-class
operators $\{\rho\}$ on the Hilbert space $\hh=\bigotimes_{k=1}^N
\hh_k$. However, dealing with infinite matrices may not be very handy. An alternative, equally complete description of any quantum state
$\rho$ of a CV system can be thus provided by suitable continuous multivariate functions, such as
one of the $s$-ordered {\em characteristic functions} \cite{barnett}
\be \chi^s_\rho (\gr\xi) = \,{\rm Tr}\,[\rho \hat D({\gr\xi})] \,e^{s\|\gr\xi\|^2/2} \; , \label{cfs} \ee with $\gr\xi\in\R^{2N}$,
$\|\cdot \|$ standing for the Euclidean norm on $\R^{2N}$. While this expression may appear quite inscrutable at first, we shall see in what follows that it can be transformed into simpler (and perhaps more familiar) functions describing the state. The
vector $\gr\xi$ belongs to the real $2N$-dimensional space
$\Gamma=({\mathbb R}^{2N},\boldsymbol{\Omega})$ equipped with a symplectic form $\boldsymbol{\Omega}$, which is called quantum {\em phase
space}, in analogy with the Liouville phase space of classical Hamiltonian mechanics. Notice that in the quantum case one cannot describe the state of a system in terms of a single phase space point (unlike classical mechanics), because of the Heisenberg uncertainty principle; phase-space regions (whose area depends on the product of the uncertainties of the canonical operators) are thus typically adopted to represent pictorially a particular state.
Observe also, from the definition of the characteristic functions, that in the
phase space picture, the tensor product structure is replaced by a
direct sum structure, so that the $N$-mode phase space decomposes as $\Gamma =
\bigoplus_k \Gamma_k$, where $\Gamma_k=({\mathbb R}^{2},\gr\omega)$ is
the local phase space associated with mode $k$.

The family of characteristic functions is in turn related, via
complex Fourier transform, to the real {\em quasi-probability
distributions} $W^s_\rho$, which constitute another set of complete
descriptions of the quantum states of a CV system \be
W^s_\rho(\gr\xi)=\frac{1}{\pi^2}\int_{{\mathbb R}^{2N}}
\chi^s_\rho(\gr\kappa) \,e^{i\gr\kappa^{\sf T} \boldsymbol{\Omega} \gr\xi}\,{\rm d}^{2N} \gr\kappa
\, . \label{qps} \ee These distributions are referred to as `quasi'-probability because they sum up to unity, yet do not behave entirely as one would expect from probability distributions. In particular, there are (infinitely many) quantum states $\rho$ for which the function $W^s_\rho$
is not a regular probability distribution for some values of $s$, as it
can assume negative values or even be singular in certain points of the phase space. An exception is the case
$s=-1$, which corresponds to the Husimi `Q-function' \cite{Husimi40}
$W^{-1}_\rho(\gr\xi)=\bra{\gr\xi}\rho\ket{\gr\xi}/\pi$ and represents a
nonnegative and regular distribution for any quantum state $\rho$. The case $s=0$ corresponds to the
so-called `Wigner function' \cite{Wigner32}, which will be discussed more extensively and denoted
simply by $W_\rho$ in the following. Likewise, for the sake of simplicity, $\chi_\rho$ will
stand for the symmetrically ordered characteristic function
$\chi^0_\rho$. Finally, the case $s=1$ yields the
so-called `P-representation', which was introduced independently by Glauber
\cite{Glauber63} and Sudarshan \cite{Sudarshan63}, and corresponds to the expansion of $\rho$ in the basis of coherent states, e.g.~for a single mode $\rho =
\int W^1_\rho(\alpha) |\alpha\rangle\langle\alpha| {\rm d}^2 \alpha$; notice that ${\rm d}^2 \alpha = {\rm d}({\rm Re}\ \alpha)\, {\rm d}({\rm Im}\ \alpha)$.

The distribution $W^1$ is the wildest of the trio, becoming negative or even singular (namely, more singular than a Dirac $\delta$) as soon as the state $\rho$ deviates from being a mixture of coherent states. For this reason, the regularity and positivity of the P-representation is often adopted as an indicator of `classicality' of a CV state $\rho$ \cite{BWallsMilburn}. Here `classical' has to be understood in the sense of  semi-classical optics, which  treats the wave completely classically but considers the quantisation of a system to take place in the detector; in this respect, even a single beam of light can be in a nonclassical state, e.g.~any Fock state with $n>0$. When dealing with multimode systems, different notions of quantumness versus classicality exist, which assess the nature of {\it correlations} among the various modes. According to such a (more informationally-oriented) view, a single system is instead always classical, and a bipartite or multipartite system is in a quantum (read, quantumly correlated) state if it violates a Bell inequality (revealing nonlocality) \cite{bellrev} or if it displays entanglement (revealing nonseparability) \cite{entanglement} or more generally if it possesses quantum discord (revealing coherence in all local bases) \cite{modireview}. The notions of classicality arising from the P-representation rather than from the nature of correlations are often in complete contrast \cite{ParisFerraro}. We will reprise these concepts later in the article.

The quasiprobability distributions of integer order $W^{-1}$, $W^0$
and $W^1$ are respectively associated with the antinormally ordered,
symmetrically ordered and normally ordered expressions of operators.
More precisely, if the operator $\hat{O}$ can be expressed as
$\hat{O}= f(\hat a_k,\hat a^{\dag}_k)$ for $k=1,\ldots,N$, where $f$
is a, say, symmetrically ordered function of the field operators,
then one has \cite{cahillglau1,cahillglau2}
\be\label{cahill}
\,{\rm Tr}[\rho \hat{O}] =
\int_{\R^{2N}}W^0_\rho(\gr\kappa)\tilde{f}(\gr\kappa) \,{\rm d}^{2N}\gr\kappa \, ,
\ee
where
$\tilde{f}(\gr\kappa)=f(\gr\kappa_k+i\gr\kappa_{k+1},\gr\kappa_k-i\gr\kappa_{k+1})$ ($k=1,\ldots,2N$)
The same relationship holds between $W_\rho^{-1}$ and the
antinormally ordered expressions of the operators and between
$W_\rho^{1}$ and the normal ordering.

Focusing on the symmetrically ordered expansions, Eq.~(\ref{cahill}) entails the following identities for the trace
\be 1 = \tr\rho = \int_{\R^{2N}} W_\rho(\gr\kappa) \,{\rm d}^{2N}\gr\kappa =
\chi_\rho(0) \,\label{wigtr} , \ee and for the purity
 \be \mu_\rho = \tr\rho^2 = (2\pi)^N \int_{\R^{2N}}
[W_\rho(\gr\kappa)]^2 \,{\rm d}^{2N}\gr\kappa = \int_{\R^{2N}} |\chi_\rho(\gr\xi)|^2
\,{\rm d}^{2N}\gr\xi    \,\label{wigpr} , \ee of a state $\rho$.
These expressions will be useful in the following.

The (symmetric) Wigner function can be written as follows in terms
of the (unnormalised) eigenvectors $\ket{\gr x}$ of the quadrature
operators $\{\hat q_j\}$ (for which $\hat q_j \ket{\gr x}=q_j \ket{\gr x}$,
$\gr x\in\rr^{N}$, for $j=1,\ldots,N$) \cite{Simon00} \be W_\rho(\gr q,\gr p) =
\frac{1}{\pi^{N}} \int_{\rr^{N}} \bra{\gr q+\gr x}\rho\ket{\gr q-\gr x}\,e^{2i\gr x\cdot \gr p} \,{\rm d}^{N}\gr x \: , \quad \gr q,\gr p\in\rr^{N}. \label{wig}
\ee From an operational point of view, the Wigner function admits a
clear interpretation in terms of homodyne measurements
\cite{sculzub}: the marginal integral of the Wigner function over
the variables $p_1,\ldots,p_N,q_1,\ldots,q_{N-1}$
\be\int_{\rr^{2N-1}}W_\rho(q_1,p_1,\ldots,q_N,p_N)\,{\rm d}\,p_1\ldots{\rm
d}\,p_{N}\,{\rm d}\,q_1\ldots{\rm
d}\,q_{N-1}=\bra{\hat{q}_N}\rho\ket{\hat{q}_N}\ee gives the correct probability distribution associated to the measurement of the remaining quadrature $\hat{q}_{N}$ (and similarly for each other quadrature)
\cite{francamentemeneinfischio}.

\section{Gaussian states and Gaussian operations}\label{secGauss}

\subsection{Gaussian states}\label{secGaussStates}

Gaussian functions are introduced early on in our learning of probability theory, often under the name of `normal distributions'. These functions appear endlessly throughout the study of probability and statistics and it would be wise for any mathematician or physicist to be familiar with them. Though perhaps not as familiar a term, Gaussian \textit{states} are analogously ubiquitous in the laboratories of quantum physicists: coherent states, such as those from a laser, which we have defined earlier; thermal states, as from a black body source, and even the vacuum state are all Gaussian. Importantly, Gaussian states are very closely related to Gaussian functions. A Gaussian state is defined as any state whose characteristic functions and quasiprobability distributions are Gaussian functions on the quantum phase space $\Gamma$. In the case of pure states, this property also coincides with a Gaussian wavefunction in the quadrature (position or momentum) basis \cite{sammyrev}.

A general multi-variate Gaussian function has the form
\begin{equation}
\label{GF}
f(\gr{x})=C\exp\left(-\frac{1}{2}\gr{x}\T\gr{A}\gr{x}+\gr{b}\T\gr{x}\right)
\end{equation}
where $\gr{x}=(x_1,x_2,...,x_N)\T$, $\gr{b}=(b_1,b_2,...,b_N)\T$, and $\gr{A}$ is an $N \times N$ positive-definite matrix.

An instructive way to start our exposition is by examining the vacuum state for a single mode $k$, introduced in Eq.~(\ref{vuoto}).
The vacuum state $|0\rangle_k$ is an eigenstate of the annihilation operator $\hat{a}_k$ with eigenvalue $0$. Expressing the annihilation operator in terms of quadratures, $\hat{a}_k=\frac{1}{\sqrt{2}}(\hat{q}_k+i\hat{p}_k)$, we can easily evaluate the vacuum wavefunction expressed in the $q$-quadrature basis, $\psi_0(q)=\langle q|0\rangle_k$. We find
\begin{align}
\hat{a}_k|0\rangle&=\frac{1}{\sqrt{2}}(\hat{q}_k+i\hat{p}_k)\int {\rm d}q|q\rangle\langle q|0\rangle_k
=\int {\rm d} q |q\rangle \left(q + \frac{\partial}{\partial q}\right)\langle q|0\rangle_k,
\end{align}
thus $\left(q + \frac{\partial}{\partial q}\right)\psi_0(q)=0$ and \be\label{vacwf}\psi_0(q)=\frac{1}{\sqrt[4]{\pi}}e^{-\frac{q^2}{2}}.\ee
In accordance with our expectations we also find the Wigner function of the vacuum state to be a Gaussian given by \be\label{wigvac}W_{\ket{0}}(q,p)=\frac{1}{\pi}e^{-q^2-p^2}.\ee This is easily checked using the expression for a single-mode pure state Wigner function, from Eq.~(\ref{wig}): $W_{\ket{\psi}}(q,p)=\frac{1}{\pi}\int_{-\infty}^\infty e^{2ipx}\overline{\psi}^(q+x)\psi(q-x){\rm d}x$.

In general, a Gaussian state is fully characterised by its \emph{first} and \emph{second} canonical moments only. The first moments $\boldsymbol{d}$ of a state $\rho$  are defined as
\begin{eqnarray}
d_{j}=\big\langle\hat{R}_{j}\big\rangle_\rho,
\end{eqnarray}
and the second moments $\boldsymbol{\sigma}$, which form the so-called  \emph{covariance matrix} $\sig = (\sigma_{ij})$, are
\begin{eqnarray}
\label{eqn:cov-mat-def}
\sigma_{ij}=\big\langle\hat{R}_{i}\hat{R}_{j}+\hat{R}_{j}\hat{R}_{i}\big\rangle_\rho -2\big\langle\hat{R}_{i}\big\rangle_\rho\big\langle\hat{R}_{j}\big\rangle_\rho.
\end{eqnarray}
Here $\langle \hat{O} \rangle_\rho \equiv \tr[\rho\,\hat{O}]$ denotes the mean of the operator $\hat{O}$ evaluated on the state $\rho$.

We note that the covariance matrix $\boldsymbol{\sigma}$ is a real, symmetric, positive definite matrix.
In the language of statistical mechanics, the elements of the covariance matrix
are the two-point truncated correlation functions between the $2N$ canonical
continuous variables. Note that our above definitions differ from many other good references for Gaussian state quantum information (e.g.~\cite{parisbook}) in the sense that strictly speaking our covariance matrix is ``twice'' the one conventionally defined by other authors; notice, e.g., how the diagonal elements in Eq.~(\ref{eqn:cov-mat-def}) contain the doubled variances of the position and momentum operators. We remark we have also chosen the natural unit convention $\hbar=1$ which again other authors choose differently \cite{pirandolareview}. It is therefore prudent to be aware of which conventions a particular article has chosen when attempting to reproduce calculations.

As anticipated, for $N$-mode Gaussian states $\rho$, the characteristic and Wigner functions for a Gaussian state have in general a Gaussian form as in Eq.~(\ref{GF}), completely determined by $\gr{d}$ and $\sig$ and given specifically by
\begin{subequations}\label{eq:wcG}
\begin{align}
\chi_\rho(\boldsymbol{\xi})&=e^{-\frac{1}{4}\boldsymbol{\xi}\T\boldsymbol{\Omega}\boldsymbol{\sigma}\boldsymbol{\Omega}\T\boldsymbol{\xi}-i\left(\boldsymbol{\Omega}\boldsymbol{d}\right)\T\boldsymbol{\xi}}, \label{eq:charG}\\
W_\rho(\boldsymbol{X})&=\frac{1}{\pi^{N}}\frac{1}{\sqrt{\mathrm{det}(\boldsymbol{\sigma})}}e^{-(\boldsymbol{X}-\boldsymbol{d})\T\boldsymbol{\sigma}^{-1}(\boldsymbol{X}-\boldsymbol{d})},\label{eq:wigG}
\end{align}
\end{subequations}
with $\gr\xi, \gr X \in \mathbb{R}^{2N}$.

\begin{figure}[t]
\centering
\includegraphics[width=11cm]{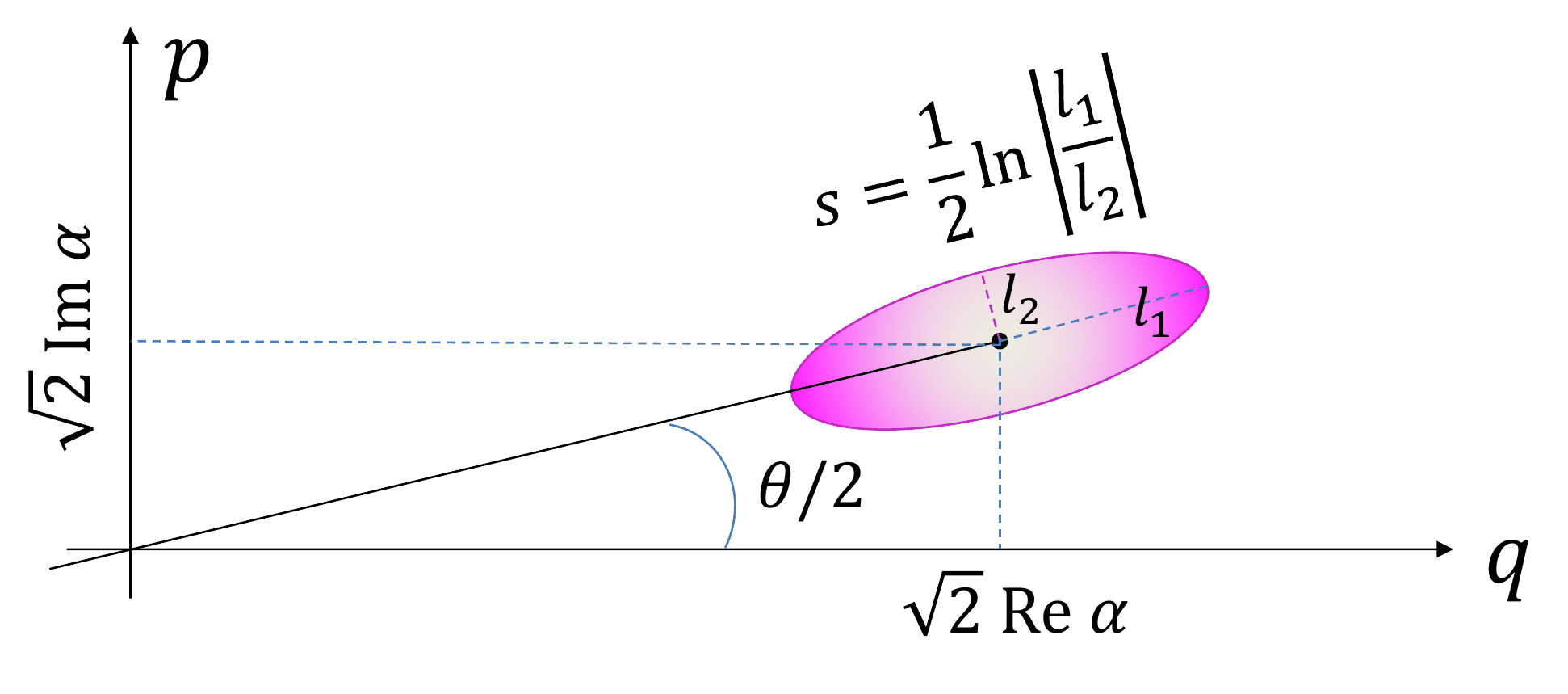}
\caption{Cross-section of the Wigner function for a general pure  Gaussian state $\ket{\psi_{\alpha,s,\theta}}_k = \hat{D}_k(\alpha) \hat{S}_k(s e^{i \theta}) \ket{0}_k$ of a single mode $k$, Eq.~(\ref{puregaussket}), characterised by a complex displacement vector $\alpha$, a real squeezing degree $s$ and a squeezing phase $\theta$. The Wigner function is given explicitly by the Gaussian form (\ref{eq:wigG}) with first moments and covariance matrix given by Eq.~(\ref{puregausscm}).
\label{figwig1}}
\end{figure}

Coherent states introduced in the previous section are instances of Gaussian states. As an example, we can compute the first moments and covariance matrix of a single-mode coherent state $\ket{\alpha}_{k}$. Using the definitions of the quadrature operators~(\ref{qpqp}) and the property $\hat{a}_{k}\ket{\alpha}_{k}=\alpha \ket{\alpha}_{k}$, we find
\begin{eqnarray}
\label{eqn:cov-mat-coherent}
\boldsymbol{d}=\sqrt{2}\left(\begin{array}{c}\mathrm{Re}(\alpha) \\ \mathrm{Im}(\alpha)\end{array}\right)\,,\quad\boldsymbol{\sigma}=\boldsymbol{I}.
\end{eqnarray}
Remarkably, the covariance matrix for a coherent state is identical for all coherent parameters $\alpha_{k}$ (which encompass the vacuum as well, $\alpha_k=0$) and is just the identity matrix.
This reflects the fact that coherent states are states of minimum Heisenberg uncertainty,
\begin{eqnarray}\label{cohunc}
\mathrm{Var}(\hat{q}_{k})\mathrm{Var}(\hat{p}_{k})=\frac{1}{4},
\end{eqnarray}
where $\mathrm{Var}_\rho(\hat{\mathcal{O}})=\langle\hat{\mathcal{O}}^{2}\rangle_\rho-\langle\hat{\mathcal{O}}\rangle_\rho^{2}$ is the variance of an observable for a given state $\rho$. Moreover, it elegantly justifies the terminology of `displacement operator', as we shall see in Fig.~\ref{figwig1}, because it effectively shifts the position of the Wigner function in phase space, whilst maintaining its shape.

Coherent states are not the only states which saturate the uncertainty relation. A larger class of states retains the property in \eq{cohunc}, but allowing for unbalanced variances on the two canonical quadratures for each mode, e.g.~a very small variance on position, and a correspondingly large uncertainty on momentum: these are called {\it squeezed states} \cite{BWallsMilburn}. The most general Gaussian pure state $\ket{\psi}_k$ of a single mode is a {\it displaced squeezed state} obtained by the combined action of the displacement operator $\hat D_k(\alpha)$ [Eq.~(\ref{CV:Weyl})] and of the (single-mode) squeezing operator $\hat{S}_k(\zeta)$,
\begin{equation}\label{CV:Sqz}
\hat{S}_k(\zeta) = \exp\left[\frac12(\zeta{\hat{a}_k^{\dagger}}{}^2-\overline{\zeta} \hat{a}_k^2)\right]\,,\quad \zeta=s e^{i \theta}\,,
\end{equation}
on the vacuum state $\ket{0}_k$:
\begin{equation}\label{puregaussket}
\ket{\psi_{\alpha,\zeta}}_k \equiv
\ket{\psi_{\alpha,s,\theta}}_k = \hat{D}_k(\alpha) \hat{S}_k(\zeta) \ket{0}_k\,,
\end{equation}

Pure single-mode Gaussian states are thus entirely specified by their displacement vector $\alpha \in \mathbb{C}$, their squeezing degree $s\in \mathbb{R}^+$, and their squeezing phase $\theta \in [0,2\pi]$. A cross-section of the Wigner function [Eq.~(\ref{wig})] for one such generic state, which shows the geometric meaning of these parameters in phase space, is plotted in Fig.~\ref{figwig1}. In the following, we will provide effective tools to calculate the first and second moments of general Gaussian states, detailing in particular the phase space description associated to unitary operations, such as those realised by displacement and squeezing, on reference states like the vacuum.

First moments for $N$-mode states can be arbitrarily adjusted by local unitary
operations, namely displacements in phase space, i.e.~applications of
the single-mode Weyl operator \eq{CV:Weyl} to locally re-center the marginal Wigner function  corresponding to each single mode. Such operations leave all
informationally relevant properties, such as entropy or any measure of correlations,
invariant. Therefore, in the following (unless otherwise stated) we
will often assume that the states under our consideration have their first moments set to $\gr{d}=\gr{0}$, without any loss of generality
for the aims of the present analysis.

Despite the infinite dimension of the associated Hilbert space,
the complete description of an arbitrary Gaussian state $\rho$ (up to
local unitary operations) is hence given by the $2N \times 2N$ covariance matrix $\sig$.
As the real $\sig$ contains the complete locally-invariant
information on a Gaussian state, it is natural to expect that  some constraints
exist which have to be obeyed by any {\em bona fide} covariance matrix, reflecting in
particular the requirement that the associated density matrix $\rho$ in Hilbert space be positive semidefinite. Indeed, such a condition together
with the canonical commutation relations imply
\begin{equation}
\sig+ i\gr\Omega\ge 0 \; , \label{bonfide}
\end{equation}
Given a Gaussian state, Inequality~(\ref{bonfide}) is the necessary and sufficient condition the matrix
$\sig$ has to fulfill to describe a physical
density matrix $\rho$ \cite{simon87,simon94}. More in general, the previous
condition is necessary for the covariance matrix of {\em any}, generally non-Gaussian, CV state (characterised in principle by nontrivial moments of any
order). We note that such a constraint implies $\sig > 0$.
Inequality~(\ref{bonfide}) is the expression of the uncertainty principle on the
canonical operators in its strong, Robertson--Schr\"odinger form
\cite{robertson30,schrodinger30,serafozziprl}.

Gaussian states $\rho$ can, of course, be pure or mixed. We can easily define pure and mixed Gaussian states by
\begin{eqnarray}
\label{eqn:gaussian-purity}
\mathrm{det}(\boldsymbol{\sig})=\left\{
  \begin{array}{l l}
    +1 \Rightarrow \mathrm{pure}\\
    > 1 \Rightarrow \mathrm{mixed}
  \end{array} \right. .
\end{eqnarray}
From Eqs.~(\ref{wigpr}) and (\ref{eq:wcG}), one has more generally that the purity of a $N$-mode Gaussian state is given by the simple formula
\begin{equation}\label{purity}
\mu_\rho = \tr\rho^2 = \frac{1}{\sqrt{\det{\sig}}}\,.
\end{equation}

A thermal equilibrium state of a single mode with frequency $\omega_k$ at temperature $T$ has covariance matrix given by \be\label{termique}
\sig = \left(\frac{\hbar \omega_k}{2 k_B T}\right) \gr{I} = (2 \overline{n}_k+1) \gr{I}\,,\ee
where the mean number of excitations (e.g.~photons, phonons) is distributed according to the Bose-Einstein statistics,
\be \overline{n}_k=\left[\exp\left(\frac{\hbar \omega_k}{k_B T}\right)-1\right]^{-1}\,.\ee
The vacuum $\overline{n}_k=0$ is retrieved for $T=0$. In general, for an arbitrary Gaussian state $\rho$ with zero first moments, the trace of the covariance matrix is directly linked to the mean energy per mode, i.e.~the average of the noninteracting Hamiltonian \eq{CV:Ham},
\begin{equation}\label{energytrace}
\overline{n}_k=\langle\hat{a}^\dagger_k \hat{a}_k\rangle_\rho = \frac14 \big[\tr (\sig) - 2\big]\,.
\end{equation}

It can be instructive to recall that, among pure states, Gaussian states are the only ones with an everywhere  positive  Wigner distribution in phase space, as proven by Hudson \cite{Hudson}. This is no longer true in the case of mixed states: mixtures of Gaussian states are, in general, non-Gaussian, yet their Wigner distribution remains obviously positive; there further exist non-Gaussian mixed states with a positive Wigner function yet such that they cannot be written as mixtures of pure Gaussian states. Recent criteria to identify non-Gaussian states with positive Wigner function have been proposed and implemented \cite{Mandilara,Filip111,Filip112,Genoni2013}.

\subsection{Gaussian unitaries and the symplectic group}

Having  defined the basic quantities that represent a Gaussian state, one may ask: \emph{how are unitary transformations represented?}
Unitary transformations on a Hilbert space are mapped to real \emph{symplectic transformations} on the first and second moments as
\begin{equation}
\label{eqn:symp-transformation}
\rho'=\hat{U}\rho\hat{U}^{\dag}\rightarrow\left\{
  \begin{array}{l l}
    \boldsymbol{d}'=\boldsymbol{S}\boldsymbol{d} \\
    \boldsymbol{\sigma}'=\boldsymbol{S}\boldsymbol{\sigma}\boldsymbol{S}\T
  \end{array} \right. ,
\end{equation}
where $\boldsymbol{S}$ is a \emph{symplectic matrix} which corresponds to the action of $\hat{U}$ on the state $\hat{\rho}$. This simple transformation rule only holds, however, for unitary transformations whose exponents are, at most, quadratic in the mode operators $\lbrace \hat{a}_{k},\hat{a}_{k}^{\dag}\rbrace$. Such unitary transformations preserve the Gaussian nature of the states. This can be intuitively understood insofar as higher order terms in the mode operators would affect higher than second-order moments. Gaussian state quantum information is built around these symplectic transformations. It will therefore be useful to introduce some of their properties.

\subsubsection{Symplectic geometry and the Williamson theorem}
Symplectic geometry has its foundations firmly rooted in classical mechanics. However, as stated earlier, it also has a place in quantum theory with profound and deep consequences. For an excellent introductory text on classical symplectic geometry see Berndt~\cite{berndt2001} and for its application to quantum mechanics see de Gosson~\cite{gosson2006}. For a particularly useful summary of symplectic geometry and its use in CV quantum systems see Arvind et. al.~\cite{Arvind95}.


The group of real symplectic matrices is defined by the condition
\begin{eqnarray}
\label{eqn:symplectic-def-1}
\boldsymbol{S}\boldsymbol{\Omega}\boldsymbol{S}\T=\boldsymbol{\Omega},
\end{eqnarray}
where $\boldsymbol{\Omega}$ is the symplectic form defined via Eq.~(\ref{eqn:real-symplectic-form}).
We denote this group by $\mathrm{Sp}(2N,\mathbb{R})$ and so define (notice that a swap $\gr S \leftrightarrow \gr S \T$ can be found in other references)
\begin{eqnarray}
\mathrm{Sp}(2N,\mathbb{R})=\left\lbrace\boldsymbol{S}| \boldsymbol{S}\boldsymbol{\Omega}\boldsymbol{S}\T=\boldsymbol{\Omega}\right\rbrace.
\end{eqnarray}
Note that symplectic matrices are always square ($2N\times 2N$), invertible matrices with determinant $\mathrm{det}(\boldsymbol{S})=+1$. Given the arrangement of operators in the basis of~(\ref{CV:R}), we decompose the symplectic matrix into the block form
\begin{eqnarray}
\boldsymbol{S}=\left(
\begin{array}{cccc}
\boldsymbol{s}_{11} & \boldsymbol{s}_{12} & \cdots & \boldsymbol{s}_{1N} \\
\boldsymbol{s}_{21} & \boldsymbol{s}_{22} &  &  \\
\vdots &  & \ddots &  \\
\boldsymbol{s}_{N1} &  &  & \boldsymbol{s}_{NN}
\end{array}
\right),
\end{eqnarray}
where the $2\times 2$ sub-block $\boldsymbol{s}_{mn}$ represents the transformation between the modes $m$ and $n$. This relates back to having unitaries whose exponents are at most quadratic in the mode operators, allowing at most pairwise mode interactions.

Williamson showed that any symmetric positive-definite matrix can be put into a diagonal form via a symplectic transformation.  An important use of this result, which amounts physically to a normal mode decomposition, is in finding the so-called symplectic eigenvalues of an arbitrary Gaussian state characterised by a covariance matrix $\boldsymbol{\sigma}$. This statement is formalised in the following theorem~\cite{williamson}:
\begin{theorem}\label{theowil}
Let $\boldsymbol{\sigma}$ be a $2N\times 2N$ positive-definite matrix. Then there exists a  $\boldsymbol{S}\in\mathrm{Sp}(2N,\mathbb{R})$ that diagonalises $\boldsymbol{\sigma}$ such that
\begin{eqnarray}
\boldsymbol{\boldsymbol{\sigma}}=\boldsymbol{S}\,\bigoplus\limits_{k=1}^{N}
\left(
\begin{array}{cc}\nu_{k} & 0 \\ 0 & \nu_{k}
\end{array}
\right)\boldsymbol{S}\T
\end{eqnarray}
\end{theorem}
A proof of this theorem can be found e.g.~in~\cite{SeralePHD,gosson2006}. We can collect the $N$ eigenvalues $\nu_{k}$ into $\boldsymbol{\nu}=\mathrm{diag}(\nu_{1},\ldots,\nu_{N})$ (either a diagonal matrix or vector). $\boldsymbol{\nu}$ is known as the \emph{symplectic spectrum} of $\boldsymbol{\sigma}$. For a physical state, the symplectic eigenvalues must be $\nu_k \geq 1$ $\forall k=1,\ldots,N$; this can be seen as equivalent to the {\it bona fide} condition (\ref{bonfide}).

 As a consequence of the Williamson decomposition, we can characterise the purity~(\ref{purity}) of a Gaussian state by rewriting its determinant as
\begin{eqnarray}
\mathrm{det}(\boldsymbol{\sigma})=\prod_{k}\nu_{k}^{2}.
\end{eqnarray}
One might use the Williamson theorem directly to find the symplectic spectrum of the covariance matrix $\boldsymbol{\sigma}$. In practice, however, it is usually much more convenient to obtain the spectrum $\boldsymbol{\nu}$ from the relation~\cite{ourreview,gosson2006}
\begin{equation}
\label{eqn:symplectic-spectrum}
\boldsymbol{\nu}=\mathrm{Eig}_{+}\left(i\boldsymbol{\Omega}\boldsymbol{\sigma}\right),
\end{equation}
where $\mathrm{Eig}_{+}\left(\boldsymbol{A}\right)$ denotes the diagonal matrix of \emph{positive} (orthogonal) eigenvalues of the matrix $\boldsymbol{A}$. The $N$ symplectic eigenvalues are thus determined by
$N$ invariants of the characteristic polynomial of the matrix  $|i\boldsymbol{\Omega}\boldsymbol{\sigma}|$
\cite{serafozziprl}.
 Knowing the symplectic spectrum of a given covariance matrix is very powerful. As we shall see, it will allow us to cast several informational measures into functions of $\boldsymbol{\nu}$; the purity above is just one example.

The {\em symplectic rank} $\aleph$ of a covariance matrix $\sig$ is defined as the number of its symplectic eigenvalues different
from $1$, corresponding to the number of normal modes which are not in the vacuum
\cite{generic}. A Gaussian state is pure if and only if $\aleph=0$.
For mixed $N$-mode states one has $1\le \aleph \le N$.
This is somehow in analogy with the standard rank of finite-dimensional
(density) matrices, defined as the number of nonvanishing
eigenvalues; in that case, pure states
$\rho=\ketbra{\psi}{\psi}$ have rank $1$, and mixed states have
in general higher rank. More specifically, for pure Gaussian states, one has $- \sig \gr\Omega \sig \gr\Omega = \gr{I}$. For generally mixed states, $\aleph(\sig) = \frac12 {\rm rank}(- \sig \gr\Omega \sig \gr\Omega - \gr{I})$. We also have, as mentioned before, that the eigenvalues of $i \gr\Omega \sig$ are $\pm \nu_i$, hence the eigenvalues of $i \gr\Omega \sig - \gr{I}$ are so formed: $0$, with degeneracy $(N-\aleph)$; $-2$, with degeneracy $(N-\aleph)$, and then   $\nu_i - 1$, $-\nu_i -1$, for all those $i=1,\ldots,\aleph$ such that $\nu_i \neq 1$.  Hence ${\rm rank}(i \gr\Omega \sig - \gr{I})=2N-(N-\aleph) = N+ \aleph$.

A visual summary of the phase space versus Hilbert space descriptions of Gaussian states, adapted from \cite{ourreview}, is offered in Table \ref{tabba}.

\begin{table}[t!]   \centering{
\begin{tabular}{ccc}
 \hline \hline
Property & Hilbert space $\hh$ & Phase space $\Gamma$ \\
\hline\hline \\
dimension & $\infty$ & $2N$ \\ & & \\ 
structure & $\bigotimes$ & $\bigoplus$ \\ & & \\ 
description & $\rho$ & $\gr d,\, \sig$ \\ & & \\ 
{\em bona fide} & $\rho \ge 0$ & $\sig + i \gr\Omega \ge 0$ \\ & & \\ 
$\begin{array}{c}\mbox{unitary} \\ \mbox{operations} \end{array}$ & $\begin{array}{c}{\hat{U}\ |\  \hat U^\dag \hat U = \hat{\id}}\\ \rho \mapsto \hat U \rho \adj{\hat U}
\end{array}$ &
$\begin{array}{c}
{\gr S}\ |\  \gr S \gr\Omega \gr S\T =
\gr\Omega \\
{\gr d \mapsto \gr S \gr d,\, \sig \mapsto \gr S \sig \gr S\T}
\end{array}$
\\ & & \\ 
spectra&
$\overset{{}}{\underset{0 \le \lambda_j \le 1}{\op{U}^{\dag} \rho \op{U} = {\rm diag}\{\lambda_j\}_{j=1}^{\infty}}}$ &
$\underset{1 \le \nu_k < \infty}{\gr S\T \sig \gr S = {\rm diag}\{(\nu_k,\nu_k)\}_{k=1}^N}$
\\ & & \\ 
pure states & $\lambda_i = 1,\,\lambda_{j \neq i}=0$ & $\nu_k = 1,\, \forall k=1,\ldots, N$
\\ & & \\ 
purity & $\overset{}{\text{Tr}{\rho^2} = \sum_j \lambda_j^2}$ &
$\overset{}{1/\sqrt{\det{\sig}}=\prod_k \nu_k^{-1}}$ \\
& & \\ \hline \hline
\end{tabular}}
\caption{Schematic comparison between Hilbert space  and  phase space  pictures
for $N$-mode Gaussian states. Note: the unitary operations $\hat{U}$ are assumed to be quadratic, as in (\ref{hatu}).}
\label{tabba}\end{table}

A natural next step would be now to link, in a concrete way, the relationship between a unitary operator and its symplectic representation. However, {\it en route} to doing so we will benefit from reviewing the different bases we can choose to write a symplectic matrix in \cite{antthesis}.

\subsubsection{Representations of $\mathrm{Sp}(2N,\mathbb{R})$}\label{sec:symplectic-representations}

An alternative representation of $\mathrm{Sp}(2N,\mathbb{R})$, from~(\ref{eqn:symplectic-def-1}), is given by the transformation
\begin{eqnarray}\label{eqn:quadrature-basis}
\hat{\gr{Y}}=\left(\begin{array}{c}
\hat{q}_{1}\\
\vdots \\
\hat{q}_{N}\\
\hat{p}_{1}\\
\vdots\\
\hat{p}_{N}
\end{array}\right)\equiv
\boldsymbol{T}
\left(\begin{array}{c}
\hat{q}_{1}\\
\hat{p}_{1}\\
\vdots\\
\vdots\\
\hat{q}_{N}\\
\hat{p}_{N}
\end{array}\right),
\end{eqnarray}
where $\boldsymbol{T}=(T_{ij})$ is a basis changing matrix with elements \be T_{ij}=\delta_{j,2i-1}+\delta_{j+2N,2i} \ee for $i,j=1,\ldots,2N$. This is, of course, nothing more than a rearrangement of the original basis~(\ref{CV:R}). We can refer to the vector $\hat{\gr{Y}}$ in Eq.~(\ref{eqn:quadrature-basis})
as the \emph{quadrature basis} vector.

The benefit of this basis is that the symplectic form and any symplectic matrix now takes a block form (with an abuse of notation calling the symplectic form and symplectic matrices with respect to this new basis again $\boldsymbol{\Omega}$ and $\boldsymbol{S}$).
\begin{eqnarray}\label{eqpasc}
\boldsymbol{\Omega}=\left(
\begin{array}{cc}
\boldsymbol{0} & \boldsymbol{I} \\
-\boldsymbol{I} & \boldsymbol{0}
\end{array}
\right),
\hspace{5mm}
\boldsymbol{S}=\left(
\begin{array}{cc}
\boldsymbol{A} & \boldsymbol{B} \\
\boldsymbol{C} & \boldsymbol{D}
\end{array}
\right).
\end{eqnarray}
Note the above blocks are now $N\times N$ dimensional. We can use the defining symplectic relation~(\ref{eqn:symplectic-def-1}) $\boldsymbol{S} \boldsymbol{\Omega} \boldsymbol{S}\T = \boldsymbol{\Omega}$ to find the following conditions\footnote{The conditions in (\ref{conpasc}) are easily proved. Using the notation in (\ref{eqpasc}), we have $$
\boldsymbol{S} \boldsymbol{\Omega} \boldsymbol{S}\T =
\left(\!\!
\begin{array}{cc}
\boldsymbol{A} & \boldsymbol{B} \\
\boldsymbol{C} & \boldsymbol{D}
\end{array}\!\!
\right)
\left(\!\!
\begin{array}{cc}
\boldsymbol{0} & \boldsymbol{I} \\
-\boldsymbol{I} & \boldsymbol{0}
\end{array}\!\!
\right)
\left(\!\!
\begin{array}{cc}
\boldsymbol{A}\T & \boldsymbol{C}\T \\
\boldsymbol{B}\T & \boldsymbol{D}\T
\end{array}\!\!
\right) =
\left(\!\!
\begin{array}{cc}
-\boldsymbol{B}\boldsymbol{A}\T+\boldsymbol{A}\boldsymbol{B}\T & -\boldsymbol{B}\boldsymbol{C}\T+\boldsymbol{A}\boldsymbol{D}\T \\
-\boldsymbol{D}\boldsymbol{A}\T+\boldsymbol{C}\boldsymbol{B}\T & -\boldsymbol{D}\boldsymbol{C}\T+\boldsymbol{C}\boldsymbol{D}\T
\end{array}\!\!
\right).$$
Equating the right-hand side to $\boldsymbol{\Omega}=\left(\!\!
\begin{array}{cc}
\boldsymbol{0} & \boldsymbol{I} \\
-\boldsymbol{I} & \boldsymbol{0}
\end{array}\!\!
\right)
$ as required by the symplectic definition~(\ref{eqn:symplectic-def-1}), we get the three independent conditions $\boldsymbol{A}\boldsymbol{B}\T=\boldsymbol{B}\boldsymbol{A}\T,\,\boldsymbol{C}\boldsymbol{D}\T=\boldsymbol{D}\boldsymbol{C}\T,\,\boldsymbol{A}\boldsymbol{D}\T-\boldsymbol{B}\boldsymbol{C}\T=\boldsymbol{I}$, which amount to Eqs.~(\ref{conpasc}).}  for $\boldsymbol{A},\boldsymbol{B},\boldsymbol{C}$ and $\boldsymbol{D}$
\begin{subequations}\label{conpasc}
\begin{align}
\left(\boldsymbol{A}\boldsymbol{B}\T\right)\T&=\boldsymbol{A}\boldsymbol{B}\T,\\
\left(\boldsymbol{C}\boldsymbol{D}\T\right)\T&=\boldsymbol{C}\boldsymbol{D}\T,\\
\boldsymbol{A}\boldsymbol{D}\T&-\boldsymbol{B}\boldsymbol{C}\T=\boldsymbol{I}.
\end{align}
\end{subequations}
The corresponding expression for the Williamson normal form is then
\begin{eqnarray}
\boldsymbol{\boldsymbol{\sigma}}=\boldsymbol{S}\left(\begin{array}{cc}\boldsymbol{\nu} & \boldsymbol{0} \\ \boldsymbol{0} & \boldsymbol{\nu} \end{array}\right)\boldsymbol{S}\T,
\end{eqnarray}
where $\boldsymbol{\nu}$ is the previously defined symplectic spectrum associated with the covariance matrix $\boldsymbol{\sigma}$.

Using this representation, we can naturally transform to a new basis which is known as the \emph{complex form} of $\mathrm{Sp}(2N,\mathbb{R})$~\cite{Arvind95}. Note that this is not a ``complexification" of the group, it is simply a change of basis which is very convenient. It is essentially a transformation from the quadrature operators $\hat{q}_{j},\hat{p}_{j}$ to the mode operators $\hat{a}_{j},\hat{a}_{j}^{\dag}$ given by
\begin{eqnarray}\label{eqn:mode-op-basis}
\hat{\boldsymbol{\xi}}=\left(\begin{array}{c}
\hat{a}_{1}\\
\vdots \\
\hat{a}_{N}\\
\hat{a}_{1}^{\dag}\\
\vdots\\
\hat{a}_{N}^{\dag}
\end{array}\right)\equiv
\boldsymbol{L}_{(c)}
\left(\begin{array}{c}
\hat{q}_{1}\\
\vdots \\
\hat{q}_{N}\\
\hat{p}_{1}\\
\vdots\\
\hat{p}_{N}
\end{array}\right),
\end{eqnarray}
where the basis changing matrix elegantly reads
\begin{eqnarray}\label{matrixL}
\boldsymbol{L}_{(c)}=\frac{1}{\sqrt{2}}
\left(\begin{array}{cc}
\boldsymbol{I} & i\boldsymbol{I} \\
\boldsymbol{I} & -i\boldsymbol{I}
\end{array}\right).
\end{eqnarray}
In this representation, we can find the complex form of any matrix written in the quadrature basis~(\ref{eqn:quadrature-basis}) via
\begin{eqnarray}
\label{eqn:complex-form-embedding}
\boldsymbol{S}\rightarrow \boldsymbol{S}_{(c)}=\boldsymbol{L}_{(c)}\boldsymbol{S}\boldsymbol{L}_{(c)}^{\dag}.
\end{eqnarray}
Using this rule, we find that the complex forms of the symplectic matrices are particularly aesthetically pleasing~\cite{Arvind95}:
\begin{eqnarray}
\label{eqn:complex-form-defs}
\boldsymbol{\Omega}_{(c)}=-i\boldsymbol{K},
\hspace{5mm}
\boldsymbol{K}=\left(
\begin{array}{cc}
\boldsymbol{I} & \boldsymbol{0} \\
\boldsymbol{0} & -\boldsymbol{I}
\end{array}
\right),
\hspace{5mm}
\boldsymbol{S}_{(c)}=\left(
\begin{array}{cc}
\boldsymbol{\alpha} & \boldsymbol{\beta} \\
\overline{\boldsymbol{\beta}} & \overline{\boldsymbol{\alpha}}
\end{array}
\right).
\end{eqnarray}
In addition, the defining symplectic relation~(\ref{eqn:symplectic-def-1}) is replaced by
\begin{eqnarray}
\label{eqn:symplectic-relation-complex}
\boldsymbol{S}_{(c)}\boldsymbol{K}\boldsymbol{S}_{(c)}^{\dag}=\boldsymbol{K},
\end{eqnarray}
where we notice that the transposition operation has been promoted to a Hermitian conjugation due to the embedding~(\ref{eqn:complex-form-embedding}). Using~(\ref{eqn:symplectic-relation-complex}), we find that the conditions for $\boldsymbol{S}_{(c)}$ to be symplectic result in the expressions
\begin{subequations}
\begin{align}
\boldsymbol{\alpha}\boldsymbol{\alpha}^{\dag}-\boldsymbol{\beta}\boldsymbol{\beta}^{\dag}&=\boldsymbol{I},\\
\boldsymbol{\alpha}\boldsymbol{\beta}\T&=\left(\boldsymbol{\alpha}\boldsymbol{\beta}\T\right)\T.
\end{align}
\end{subequations}
Finally, the Williamson normal form for the complex form of $\mathrm{Sp}(2N,\mathbb{R})$ reads
\begin{eqnarray}
\boldsymbol{\boldsymbol{\sigma}}_{(c)}=\boldsymbol{S}_{(c)}\left(\begin{array}{cc}\boldsymbol{\nu} & \boldsymbol{0} \\ \boldsymbol{0} & \boldsymbol{\nu} \end{array}\right)\boldsymbol{S}^{\dag}_{(c)},
\end{eqnarray}
where $\boldsymbol{\nu}$ remains unchanged from the previous definitions and $\boldsymbol{\boldsymbol{\sigma}}_{(c)}$ is the complex form of the covariance matrix $\boldsymbol{\boldsymbol{\sigma}}$, in the basis of mode operators defined via
\begin{eqnarray}
(\boldsymbol{\boldsymbol{\sigma}}_{(c)})_{mn}=\langle\hat{\xi}_{m}\hat{\xi}_{n}^{\dag}+\hat{\xi}_{m}^{\dag}\hat{\xi}_{n}\rangle-2\langle\hat{\xi}_{n} \rangle\langle\hat{\xi}_{m}^{\dag}\rangle.
\end{eqnarray}
It should be noted that in the complex form, the symplectic spectrum~(\ref{eqn:symplectic-spectrum}) of a covariance matrix can be computed using
\begin{eqnarray}
\label{eqn:symplectic-spectrum-complex}
\boldsymbol{\nu}=\mathrm{Eig}_{+}(\boldsymbol{K}\boldsymbol{\sigma}_{(c)}).
\end{eqnarray}
Of course, the complex form covariance matrix $\boldsymbol{\boldsymbol{\sigma}}_{(c)}$ can be obtained from $\boldsymbol{\boldsymbol{\sigma}}$ by the transformation rule~(\ref{eqn:complex-form-embedding}) which results in the block form
\begin{eqnarray}
\boldsymbol{\boldsymbol{\sigma}}_{(c)}=\left(\begin{array}{cc}
\mathbfcal{V} & \mathbfcal{U} \\
\overline{\mathbfcal{U}} & \overline{\mathbfcal{V}}
\end{array} \right),
\end{eqnarray}
with the conditions $\mathbfcal{V}^{\dag}=\mathbfcal{V}$ and $\mathbfcal{U}\T=\mathbfcal{U}$. This implies that $\boldsymbol{\boldsymbol{\sigma}}_{(c)}^{\dag}=\boldsymbol{\boldsymbol{\sigma}}_{(c)}$.

 The Lie algebra of the symplectic group will help us derive equations which govern the time evolution of a quantum state  (see e.g. Hall~\cite{hall2004} for a reference on Lie groups and algebras).
To begin, we define a set of Hermitian, $2N\times 2N$, linearly independent basis matrices $\boldsymbol{G}_{j}$, such that
\begin{eqnarray}
\mathfrak{sp}(2N,\mathbb{R})=\left\lbrace \boldsymbol{K}\boldsymbol{G}_{j}| \boldsymbol{G}_{j}^{\dag}=\boldsymbol{G}_{j}\right\rbrace .
\end{eqnarray}
$\mathfrak{sp}(2N,\mathbb{R})$ is known as the \emph{Lie algebra} of $\mathrm{Sp}(2N,\mathbb{R})$. We can link a Lie algebra with its group via the exponential map~\cite{hall2004}. The real symplectic group is \emph{connected} (though not simply connected) and is \emph{non-compact}. Being non-compact implies that not every symplectic matrix can be written as the exponential of a single matrix. However, as we will see, every element of the symplectic group \emph{can} be written as a product of exponentials.

The matrices $\boldsymbol{K}\boldsymbol{G}_{j}$ are the infinitesimal generators of $\mathrm{Sp}(2N,\mathbb{R})$ and form a finite, closed algebra of dimension $N(2N+1)$. To ensure the correct properties of the symplectic group, the matrices $\boldsymbol{G}$ are necessarily of the form
\begin{eqnarray}
\boldsymbol{G}=\left(\begin{array}{cc}
\mathbfcal{X} & \mathbfcal{Y} \\
\overline{\mathbfcal{Y}} & \overline{\mathbfcal{X}}
\end{array} \right),
\end{eqnarray}
with the conditions $\mathbfcal{X}^{\dag}=\mathbfcal{X}$ and $\mathbfcal{Y}\T=\mathbfcal{Y}$. Note that the matrices $\boldsymbol{G}_{j}$ are not entirely arbitrary  and have dimension $\mathrm{dim}(\mathbfcal{X})+\mathrm{dim}(\mathbfcal{Y})=N^{2}+2N+N(N-1)=N(2N+1)$ as stated earlier (a general  $2N\times 2N$ Hermitian matrix has dimension $4N^{2}$). A useful consequence of the algebra being closed is that we can decompose any symplectic matrix in the product decomposition
\begin{eqnarray}
\boldsymbol{S}_{(c)}=\prod_{j}e^{-ig_{j}\boldsymbol{K}\boldsymbol{G}_{j}},
\end{eqnarray}
where $g_{j}\in\mathbb{R}$ and the product runs over the $N(2N+1)$ independent symplectic generators.   The complex form of $\mathrm{Sp}(2N,\mathbb{R})$ allows us to construct the form of $\gr{S}$ associated to a quadratic unitary $\hat{U}$.

\subsubsection{Symplectic representation of linear optics operations}
The backbone of {\it linear optics}  is constituted by unitary transformations whose exponent is quadratic in the field operators \cite{BWallsMilburn}. The pervasiveness of linear optics can be understood by noting the large energies required to achieve high order non-linear effects on single-modes, and the similar difficulties encountered when interacting more than two modes at at time.
Explicitly, \be\hat{U}=e^{-i \hat{H}}\,,\label{hatu}\ee with a generic quadratic Hamiltonian (we omit linear terms as they can be reabsorbed by local displacements as mentioned before)
\begin{eqnarray}\label{isthatham}
\hat{H}=A_{mn}\hat{a}^{\dag}_{m}\hat{a}_{n}+B_{mn}\hat{a}^{\dag}_{m}\hat{a}^{\dag}_{n}
+\overline{B}_{mn}\hat{a}_{m}\hat{a}_{n}+\overline{A}_{mn}\hat{a}_{m}\hat{a}_{n}^{\dag}.
\end{eqnarray}

Transformations of this form take an arbitrary linear combination of field operators to another arbitrary linear combination of field operators. Mathematically we have, for complex coefficients $\alpha_{jk},\beta_{jk}$,
\begin{equation}
\begin{aligned}
\hat{U}\hat{a}_{k}\hat{U}^{\dag}&=&\sum_{j}\alpha_{jk}\hat{a}_{j}+\sum_{j}\beta_{jk}\hat{a}_{j}^{\dag},\\
\hat{U}\hat{a}_{k}^{\dag}\hat{U}^{\dag}&=&\sum_{j}\overline{\alpha}_{jk}\hat{a}_{j}^{\dag}+\sum_{j}\overline{\beta}_{jk}\hat{a}_{j},
\end{aligned}
\end{equation}
which can be written compactly as
\begin{eqnarray}
\hat{U}\left(\begin{array}{c}\hat{\boldsymbol{a}}\\ \hat{\boldsymbol{a}}^{\dag}\end{array}\right)\hat{U}^{\dag}=\left(\begin{array}{cc} \boldsymbol{\alpha} & \boldsymbol{\beta} \\ \overline{\boldsymbol{\beta}} & \overline{\boldsymbol{\alpha}} \end{array}\right) \left(\begin{array}{c} \hat{\boldsymbol{a}}\\ \hat{\boldsymbol{a}}^{\dag}\end{array}\right).
\end{eqnarray}
As these linear transformations (so-called Bogolubov transformations)  must preserve the commutation relations, due to $\hat{U}$ being unitary, we find the conditions
\begin{subequations}
\label{eqn:symplectic-def-bogo}
\begin{align}
\boldsymbol{\alpha}\boldsymbol{\alpha}^{\dag}-\boldsymbol{\beta}\boldsymbol{\beta}^{\dag}&=\boldsymbol{I},\\
\boldsymbol{\alpha}\boldsymbol{\beta}\T&=\left(\boldsymbol{\alpha}\boldsymbol{\beta}\T\right)\T.
\end{align}
\end{subequations}
Remarkably, the conditions on $\boldsymbol{\alpha}$ and $\boldsymbol{\beta}$ are nothing more than the defining relations for the complex form of $\mathrm{Sp}(2N,\mathbb{R})$. Thus we can write
\begin{eqnarray}
\gr{S}_{(c)}=\left(\begin{array}{cc} \boldsymbol{\alpha} & \boldsymbol{\beta} \\ \overline{\boldsymbol{\beta}} & \overline{\boldsymbol{\alpha}} \end{array}\right).
\end{eqnarray}
The correspondence between unitary transformations (or linear transformations) and symplectic relations, i.e.
\begin{eqnarray}
\hat{U}\hat{\boldsymbol{\xi}}\hat{U}^{\dag}=\gr{S}_{(c)}\cdot\hat{\boldsymbol{\xi}},
\end{eqnarray}
allows us to use the power of symplectic geometry to calculate linear transformations of our systems. In particular, it can be shown that given a unitary operator which can be written as a single exponential (i.e.~the time ordered integral can be performed trivially), the corresponding symplectic matrix can be written as single exponential also, as shown in the following.

Recall that the commutation relations for the mode operators arranged in the vector~(\ref{eqn:mode-op-basis}) can be written compactly as
\begin{eqnarray}
\left[\hat{\xi}_{m},\hat{\xi}_{n}^{\dag}\right]=K_{mn},
\end{eqnarray}
where $K_{mn}$ are the components of $\boldsymbol{K}$ defined in~(\ref{eqn:complex-form-defs}). A generic quadratic Hamiltonian~(\ref{isthatham}) can be then  written as
\begin{eqnarray}
\label{eqn:quadraticoperator}
\hat{H}=\hat{\boldsymbol{\xi}}^{\dag}\cdot\gr{H}\cdot\hat{\boldsymbol{\xi}},
\end{eqnarray}
where the matrix representation of $\hat{H}$ takes the form
\begin{eqnarray}\label{HAB}
\boldsymbol{H}=\left(\begin{array}{cc}
\boldsymbol{A} & \boldsymbol{B} \\
\overline{\boldsymbol{B}} & \overline{\boldsymbol{A}}
\end{array}\right),
\end{eqnarray}
with the specific conditions $\boldsymbol{A}=\boldsymbol{A}^{\dag}$ and $\boldsymbol{B}=\boldsymbol{B}\T$. The conditions on $\boldsymbol{A}$ and $\boldsymbol{B}$ ensure the Hermiticity of $\boldsymbol{H}$. Next consider the unitary transformation $\hat U$ on the vector of operators $\hat{\boldsymbol{\xi}}$ such that
\begin{eqnarray}
\label{eqn:unitary-transformation}
e^{-i\hat{H}}\hat{\xi}_{m}e^{+i\hat{H}}=\sum_n S_{mn}\hat{\xi}_{n},
\end{eqnarray}
where $S_{mn}$ will be identified with a symplectic matrix (in the complex representation). One could use the Hadamard lemma
\begin{eqnarray}
e^{-\hat{X}}\hat{Y}e^{+\hat{X}}=\hat{Y}-[\hat{X},\hat{Y}]+\frac{1}{2!}[\hat{X},[\hat{X},\hat{Y}]]+\ldots
\end{eqnarray}
to find the explicit form for each transformation, however using the commutation relations of our bosonic operators we can find a link between a given quadratic unitary operator and its symplectic counterpart. In terms of the operators in~(\ref{isthatham}) it is straightforward to show, after a bit of commutator algebra, that Eq.~(\ref{eqn:unitary-transformation}) can be written as
\begin{eqnarray}
\left[-i\hat{H},\hat{a}_{k}\right]&=&-i\left(A_{km}\hat a_{m}+B_{km}\hat a_{m}^{\dag}\right),\\
\left[-i\hat{H},\hat{a}_{k}^{\dag}\right]&=&+i\left(\overline{B}_{km}\hat a_{m}^{\dag}+\overline{A}_{km}\hat a_{m}\right),
\end{eqnarray}
which is conveniently expressed in matrix form as
\begin{eqnarray}
[-i\hat{H},\hat{\boldsymbol{\xi}}]=-i\boldsymbol{K}\boldsymbol{H}\hat{\boldsymbol{\xi}}.
\end{eqnarray}
The Hadamard lemma then gives
\begin{eqnarray}
e^{-i\hat{H}}\hat{\boldsymbol{\xi}}e^{+i\hat{H}}=e^{-i\boldsymbol{K}\boldsymbol{H}}\hat{\boldsymbol{\xi}},
\end{eqnarray}
and hence we can identify a unique correspondence\footnote{Unique up to a sign in the definition of the symplectic matrix. This  is due to the fact that the unitary group can be associated with a double covering of the symplectic group, known as the {\it metaplectic} group \cite{Arvind95}.} between  a quadratic unitary operator and a symplectic matrix as \cite{luis1995}
\begin{eqnarray}\label{eqn:sym-from-h}
\hat{U}=e^{-i\hat{\boldsymbol{\xi}}^{\dag}\cdot\boldsymbol{H}\cdot\hat{\boldsymbol{\xi}}}\rightarrow\boldsymbol{S}_{(c)}=e^{-i\boldsymbol{K}\boldsymbol{H}}.
\end{eqnarray}

\subsubsection{Passive and active transformations}

In general, symplectic transformations can be divided into {\it passive} and {\it active} ones\footnote{Notice that this is different from the notion of passive versus active transformation of coordinate systems.}. If we return to the real representation in the basis (\ref{CV:R}), any $\gr S$ is generated by exponentiation of matrices written as $\gr J\gr\Omega$, where $\gr J$ is
antisymmetric \cite{Arvind95,serafozzinazi}. Such generators can be symmetric or
antisymmetric. The transformations generated by
antisymmetric operators (i.e., with $\gr{B}=\gr{0}$ in \eq{HAB}), are orthogonal and form the compact subgroup ${\rm K}(N)= {\rm Sp}(2N,\R)\cap {\rm SO}(2N)$ of $\sy{2N}$.  Acting by congruence on
the covariance matrix $\sig$, they preserve the value of $\tr{\sig}$, i.e., the mean energy of the system, see Eq.~(\ref{energytrace}). For this reason such transformations are known as {\it passive}; relevant examples include  phase shifters and beam splitters.
On the other hand, symplectic transformations generated by symmetric operators, i.e.~by the $\gr{B}$ terms in \eq{HAB} are not orthogonal and do not preserve the energy of the system; these transformations, featuring prominently single-mode and two-mode squeezing, are thus referred to as {\it active} in the conventional language of linear optics.

We will now give examples of standard symplectic matrices $\gr{S}$ associated to relevant linear optics transformations. These can be evaluated via the correspondence in \eq{eqn:sym-from-h}.

\paragraph{Phase shift.} A single-mode rotation by an angle $\varphi/2$ in phase space, also known as phase shift, is the simplest example of a passive transformation. Its unitary form is \be \hat{U}=\exp\left(i \varphi \hat{a}^\dagger_k \hat{a}_k\right) \ee for a mode $k$. This corresponds to a quadratic generator with (complex) matrix representation $\gr{H}=-\frac{\varphi}{2} \gr{I}$ (in this case, $\gr{H}$ turns out to be real). In the real basis (\ref{CV:R}), the symplectic transformation $\gr{R}(\varphi)$ associated to a rotation can be obtained by recalling Eqs.~(\ref{eqn:quadrature-basis},\ref{matrixL},\ref{eqn:complex-form-defs},\ref{eqn:sym-from-h}). We have
\begin{eqnarray}\label{symplyrot}
\gr{R}(\varphi) &=& \gr{T}\T\boldsymbol{L}_{(c)}^\dagger e^{-i\boldsymbol{K}\boldsymbol{H}} \boldsymbol{L}_{(c)} \gr{T} \nonumber \\
&=& {\small{\left(
      \begin{array}{cc}
        1 & 0 \\
        0 & 1 \\
      \end{array}
    \right) \left(
\begin{array}{cc}
 \frac{1}{\sqrt{2}} & \frac{1}{\sqrt{2}} \\
 -\frac{i}{\sqrt{2}} & \frac{i}{\sqrt{2}} \\
\end{array}
\right) \left(
\begin{array}{cc}
 e^{\frac{i \varphi }{2}} & 0 \\
 0 & e^{-\frac{i \varphi}{2}} \\
\end{array}
\right)\left(
\begin{array}{cc}
 \frac{1}{\sqrt{2}} & \frac{i}{\sqrt{2}} \\
 \frac{1}{\sqrt{2}} & -\frac{i}{\sqrt{2}} \\
\end{array}
\right)
\left(
      \begin{array}{cc}
        1 & 0 \\
        0 & 1 \\
      \end{array}
    \right)}}\nonumber \\
    &=& \left(
\begin{array}{cc}
 \cos \left(\frac{\varphi }{2}\right) & -\sin \left(\frac{\varphi }{2}\right) \\
 \sin \left(\frac{\varphi }{2}\right) & \cos \left(\frac{\varphi }{2}\right) \\
\end{array}
\right)\,.
\end{eqnarray}

\paragraph{Single-mode squeezing.} The single-mode squeezing operator is a prototypical active transformation, described by the unitary operator
$\hat{S}_k\big(s e^{i \theta}\big)$  introduced in Eq.~(\ref{CV:Sqz}). In this case, referring to \eq{HAB}, we have $\gr{A}=(0)$ and $\gr{B}=(i s e^{i \theta})$. Adopting the same procedure as before, we obtain the symplectic representation of squeezing,
\begin{equation}\label{symplysqz1}
\gr{S}^{(1)}(s,\theta) = \left(
\begin{array}{cc}
 \cosh (s)+\cos (\theta ) \sinh (s) & \sin (\theta ) \sinh (s) \\
 \sin (\theta ) \sinh (s) & \cosh (s)-\cos (\theta ) \sinh (s) \\
\end{array}
\right)\,,
\end{equation}
which reduces to \be\gr{S}^{(1)}(s,0)={\rm diag}(e^s,e^{-s})\ee for $\theta=0$. In the latter case, this operation (for $s>0$) squeezes the momentum, reducing its variance exponentially, while correspondingly enlarging the one on position. The complementary case $\theta=\pi/2$ amounts to a squeeze of the position quadrature and a corresponding increase on the variance of the momentum quadrature.

We can now write the phase space representation of the most general pure single-mode Gaussian state $\ket{\psi}_k$. From Eq.~(\ref{puregaussket}), and following the mapping (\ref{eqn:symp-transformation}), we just need to apply the operation in Eq.~(\ref{symplysqz1}) to the vacuum state, followed by a displacement. Recall from  \eq{eqn:cov-mat-coherent} that the vacuum has covariance matrix equal to the identity, and vanishing first moments. The first and second moments of a general pure single-mode Gaussian state $\ket{\psi}_k$ are then given by
\begin{subequations}\label{puregausscm}
\begin{align}
\boldsymbol{d}&=\sqrt{2}\left(\begin{array}{c}\mathrm{Re}(\alpha) \\ \mathrm{Im}(\alpha)\end{array}\right)\,,\\
\sig&=\nonumber \gr{S}^{(1)}(s,\theta) \gr{I} {\gr{S}^{(1)}}\T(s,\theta) \\ &= \left(
\begin{array}{cc}
 \cosh (2 s)+\cos (\theta ) \sinh (2 s) & \sin (\theta ) \sinh (2 s) \\
 \sin (\theta ) \sinh (2 s) & \cosh (2 s)-\cos (\theta ) \sinh (2 s) \\
\end{array}
\right).
\end{align}
\end{subequations}
Inserting this expression into Eq.~(\ref{eq:wigG}) one obtains the Wigner function whose cross-section has been depicted in Fig.~\ref{figwig1}.

\paragraph{Beam splitter.}
Another common unitary operation is the ideal (phase-free) {\em
beam splitter}, whose action $\hat{B}_{i,j}$ on a pair of modes $i$
and $j$ is defined as
\begin{equation}\label{bsplit}
\hat{B}_{i,j}(\phi):\left\{
\begin{array}{l}
\hat a_i \rightarrow \hat a_i \cos\phi + \hat a_j\sin\phi \\
\hat a_j \rightarrow \hat a_i \sin\phi - \hat a_j\cos\phi \\
\end{array} \right.\,.
\end{equation}
A beam
splitter with transmissivity $\tau$ is a passive transformation corresponding to a rotation of
$\phi = \arccos\sqrt{\tau}$ in phase space; in particular, $\phi=\pi/4$
corresponds to a balanced 50:50 beam splitter, $\tau=1/2$. Applying the machinery introduced above, one finds that the beam splitter is described by
a symplectic transformation
\begin{equation}\label{bbs}
\gr{B}_{i,j}(\tau)=\left(
\begin{array}{cccc}
 \sqrt{\tau } & 0 & \sqrt{1 - \tau } & 0 \\
 0 & \sqrt{\tau } & 0 & \sqrt{1 - \tau } \\
 \sqrt{1 - \tau } & 0 & - \sqrt{\tau } & 0 \\
 0 & \sqrt{1 - \tau } & 0 & - \sqrt{\tau }
\end{array}
\right)\,.
\end{equation}

\paragraph{Two-mode squeezing.}
We close this gallery with the two-mode squeezing operation, an active transformation which models the physics of optical parametric amplifiers (see e.g.~\cite{francamentemeneinfischio}) and is routinely employed to create CV entanglement (see Section~\ref{secGaussEnt}). Acting on the pair of modes $i$ and $j$ via the unitary
\begin{equation}\label{tmsU}
\hat{U}_{i,j}(r) = \exp[r (\hat a_{i}^\dagger \hat a_{j}^\dagger - \hat a_{i} \hat a_{j})],
\end{equation}
it corresponds
to the symplectic transformation
\begin{equation}\label{tmsS}
\gr S^{(2)}_{i,j}(r)=\left(\begin{array}{cccc}
\cosh r&0&\sinh r&0\\
0&\cosh r&0&-\sinh r\\
\sinh r&0&\cosh r&0\\
0&-\sinh r&0&\cosh r
\end{array}\right)\, .
\end{equation}

\subsubsection{Symplectic decompositions}
The most general (mixed) Gaussian state can be obtained by applying a generic quadratic unitary, corresponding to a $N$-mode symplectic transformation $\gr S$, on a product state of single-mode thermal states, \be\rho^{_\otimes}=\bigotimes_{k=1}^N
\frac{2}{\nu_{k}+1}\sum_{n=0}^{\infty}\left(
\frac{\nu_{k}-1}{\nu_{k}+1}\right)^n\ket{n}_{k}{}_{k}\bra{n},\ee possibly followed by displacements if needed. This fact stems immediately from Williamson's Theorem~\ref{theowil}, by noticing that the state $\rho^{\otimes}$ has covariance matrix $\bigoplus_{k=1}^{N}\left(\begin{array}{cc}
\nu_k&0\\
0&\nu_k
\end{array}\right)$, where $\overline n_k = \frac{\nu_k-1}{2}$ is the mean particle number in each mode $k$, see Eq.~(\ref{termique}).

By the structure of the symplectic group, it follows that the most general $\gr S$ can be decomposed in terms of products of symplectic transformations acting on single modes or on pairs of modes only. However, an alternative decomposition, referred to as the  `Euler' or `Bloch-Messiah' decomposition of a general symplectic transformation $\gr S$, is particularly insightful. We have
\cite{Arvind95,braunsqueezirreducibile}
\begin{equation}
\gr S = \gr O \gr Z \gr O' ,\label{euler1}
\end{equation}
where $\gr O, \gr O' \in {\rm K}(N)= {\rm Sp}(2N,\R)\cap {\rm SO}(2N)$ are orthogonal
symplectic transformations, while
\be
\gr Z=\bigoplus_{j=1}^{N}\left(\begin{array}{cc}
z_j & 0 \\
0 & \frac{1}{z_j}
\end{array}\right) \; ,
\ee
with $z_{j}\ge 1$ $\forall$ $j$. The set of such $Z$'s forms a
non-compact subgroup of ${\rm Sp}(2N,\R)$ comprised of local
(single-mode) squeezing operations, see Eq.~(\ref{symplysqz1}).

The Euler decomposition nicely isolates the active and passive components of a generic $\gr S$.
When $\gr S$ acts on the vacuum ($\sig=\gr{I}$), the orthogonal matrix $\gr{O}'$ is reabsorbed and plays no role ($\gr{O}' {\gr{O}'}\T = \gr{I}$); this implies that a product of single-mode squeezings followed by a passive transformation $\gr O$ suffices to generate the covariance matrix of any $N$-mode pure Gaussian state. For instance, for the single-mode case  in Eq.~(\ref{puregausscm}), the same general state can be obtained by applying to the vacuum, in sequence, an unrotated squeezing $\gr{S}^{(1)}(s,0)$ (the active part $\gr Z$), a rotation $\gr{R}(\theta)$ (the passive part $\gr O$), and finally the displacement. Precisely, \be\sig = \gr{S}^{(1)}(s,\theta)  {\gr{S}^{(1)}}\T(s,\theta) \equiv \gr{R}(\theta)\gr{S}^{(1)}(s,0)  {\gr{S}^{(1)}}\T(s,0) \gr{R}\T(\theta)\,.\ee

\subsection{Partial tracing}
After such a detailed discussion of unitary Gaussian operations, we move on to nonunitary ones. A general treatment of Gaussian decoherence channels can be found e.g.~in \cite{serafozzijob05}, see also \cite{wolfeis,holevonew} for a more recent classification of Gaussian channels.

For future convenience, let us define and write down the first moments $\gr{d}_{1,\ldots,N}$ and covariance matrix
$\sig_{1,\ldots, N}$ of an $N$-mode Gaussian state in the real basis (\ref{CV:R}) in terms of two-dimensional subblocks as
\be
\gr{d}_{1, \ldots, N} = \left(
\begin{array}{c}
\gr{d}_1\\
\gr{d}_2\\
\vdots\\
\gr{d}_N
\end{array}
\right)
,\quad
\sig_{1,\ldots, N} = \left(\begin{array}{cccc}
\sig_{1} & \eps_{1,2}\; & \cdots & \eps_{1,N} \\
\eps_{1,2}^{\sf T}\; & \ddots & \ddots & \vdots \\
\vdots & \ddots & \ddots & \eps_{N-1,N} \\
\eps_{1,N}^{\sf T}& \cdots & \eps_{N-1,N}^{\sf T} & \sig_{N} \\
\end{array}\right) \; . \label{CM}
\ee Each subvector $\gr{d}_k$ and diagonal block $\sig_k$ correspond respectively to the first moments and the local covariance matrix
for the reduced state of mode $k$, for all
$k=1,\ldots,N$. On the other hand, the off-diagonal blocks
$\eps_{i,j}$ of the covariance matrix encode the intermodal correlations (quantum and
classical) between subsystems $i$ and $j$. The matrices $\eps_{i,j}$
all vanish for a product state.

In general, partial tracing is very easy to do at the phase space level. The covariance matrix for a reduced state of a subset of modes is obtained by just removing the entries (block rows and columns) pertaining to the excluded modes; and similarly for the displacement vectors.
For instance, the marginal state of modes $1$ and $2$, obtained by tracing over modes $3,\ldots,N$ from a generic state described by Eq.~(\ref{CM}), is characterised by
\be
\gr{d}_{1,2} =
 \left(
\begin{array}{c}
\gr{d}_1\\
\gr{d}_2
\end{array}
\right)
,\quad
\sig_{1,2} = \left(\begin{array}{cc}
\sig_{1} & \eps_{1,2}\; \\
\eps_{1,2}^{\sf T}\; & \sig_2
\end{array}\right) \; . \label{CM2}
\ee
The Gaussian partial trace can naturally be extended to more modes and can be performed as many times as needed. It is, of course, a completely positive, trace preserving map like its Hilbert space counterpart and also preserves the Gaussian nature of the states.

A general bipartite $N$-mode Gaussian state, with subsystem $A$ encompassing $N_A$ modes and subsystem $B$ encompassing $N_B=N-N_A$ modes, will be described by a covariance matrix in block form
\begin{equation}\label{eq:cms}
\sig_{AB} = \left(\begin{array}{cc}
\sig_A & {\eps}_{AB} \\
{\eps}_{AB}^{\sf T} & \sig_B\,
\end{array}\right).
\end{equation}

\subsection{Gaussian measurements}

In quantum mechanics, two main types of measurement processes are usually considered \cite{BNielsChuang}. The first type is constituted by projective (von Neumann) measurements, which are defined by a set of Hermitian positive operators $\{{\hat{\Pi}}_i\}$ such that $\sum_i {\hat{\Pi}}_i=\hat{\mathbb{I}}$ and ${\hat{\Pi}}_i {\hat{\Pi}}_j=\delta_{ij}{\hat{\Pi}}_i$. A projective measurement maps a state $\rho$ into a state \be\rho_i=\frac{{\hat{\Pi}}_i \rho {\hat{\Pi}}_i}{\text{Tr}\{{\hat{\Pi}}_i \rho {\hat{\Pi}}_i\}}\ee with probability $p_i={\text{Tr}\{{\hat{\Pi}}_i \rho {\hat{\Pi}}_i\}}$. If we focus on a local projective measurement on the subsystem $B$ of a
bipartite state $\rho_{AB}$, say ${\hat{\Pi}}_i=\mathbb{I}_A\otimes {\hat{\Pi}}_{iB}$, the subsystem $A$ is then mapped into  the conditional state \be\rho_{A|{\hat{\Pi}}_i}=\text{Tr}_B\frac{{\hat{\Pi}}_i \rho_{AB} {\hat{\Pi}}_i}{\text{Tr}\{{\hat{\Pi}}_i \rho_{AB} {\hat{\Pi}}_i\}}.\ee

The second type of quantum measurements are known as POVM (positive operator-valued measure) measurements and amount to a more general class compared to projective measurements. They are defined again in terms of a set of Hermitian positive operators $\{{\hat{\Pi}}_i\}$ such that $\sum_i {\hat{\Pi}}_i=\hat{\mathbbm{I}}$, but they need not be orthogonal in this case. In the following, by `measurement' we will refer in general to a POVM.

 In the CV case, the measurement operations mapping Gaussian states into Gaussian states are called Gaussian measurements. They can be realised experimentally by appending ancillae initialised in Gaussian states, implementing Gaussian unitary (symplectic) operations on the system and ancillary modes, and then measuring quadrature operators, which can be achieved e.g.~by means of balanced homodyne detection in the optics framework \cite{mistagauss}. Given a bipartite Gaussian state $\rho_{AB}$, any such  measurement on, say,  the $N_B$-mode subsystem $B=(B_1\ldots B_{N_B})$, is described by a POVM  of the form \cite{fiurasek07}
\begin{equation}\label{gpovm}
{\hat{\Pi}}_B(\gr{\eta}) = \pi^{-N_B} \left[\prod_{j=1}^{N_B}  \hat{D}_{B_j}(\eta_j)\right] \Lambda^{\hat{\Pi}}_B \left[\prod_{j=1}^{N_B}\hat{D}^\dagger_{B_j}(\eta_j)\right]\,,
 \end{equation}
 where
 \begin{equation}\label{weyl}
 \hat{D}_B(\eta_j) = \exp(\eta_j \hat{b}_j^\dagger - \overline{\eta}_j \hat{b}_j)
  \end{equation} is the Weyl operator (\ref{CV:Weyl}),
  $\hat{b}_j$ is the annihilation operator of the $j$-th mode of the subsystem $B$, $\pi^{-N_B}\int {\hat{\Pi}}_B(\gr{\eta}) {\rm d}^{2N_B}\gr{\eta}  = \hat{\id}$, and $\Lambda^{\hat{\Pi}}_B$ is the density matrix of a (generally mixed) $N_B$-mode Gaussian state with covariance matrix $\gr{\Gamma}_B^{\hat{\Pi}}$ which denotes the so-called seed of the measurement. The conditional state $\rho_{A|\gr{\eta}}$ of subsystem $A$, after the measurement ${\hat{\Pi}}_B(\gr{\eta})$ has been performed on $B$, has a covariance matrix $\tilde{\gr\sigma}^{{\hat{\Pi}}}_A$ independent of the outcome $\gr{\eta}$ and given by the Schur complement \cite{nogo1,nogo2,nogo3}
\begin{equation}\label{eq:schur}
\tilde{\gr\sigma}^{{\hat{\Pi}}}_A = \sig_A-\eps_{AB} (\gr\sigma_B+ \gr\Gamma_B^{\hat{\Pi}})^{-1} \eps_{AB}^{\sf T}\,,
 \end{equation}
 where the original bipartite covariance matrix $\sig_{AB}$ of the original $N$-mode  state $\rho_{AB}$ has been written in block form as in Eq.~(\ref{eq:cms}), with $\eps_{AB}$ denoting the intermodal correlations.

 \section{Measures of correlations for Gaussian states}\label{secGaussEnt}

\subsection{EPR correlations}
Historically, CV entanglement was the first form of entanglement ever defined. In their influential 1935 paper \cite{EPR35}, Einstein, Podolsky and Rosen (EPR) were describing (actually using its counterintuitive features as an argued testament against the completeness of the quantum description of the world) an ideal state which is the simultaneous eigenstate of relative position and total momentum (or vice versa) of two particles, i.e.
\begin{equation}\label{eprstate}
\ket{\psi}^{{\rm EPR}}_{AB} \sim \delta (\hat{q}_A-\hat{q}_B)\ \delta(\hat{p}_A+\hat{p}_B)\,.
\end{equation}
Such a state is in fact unphysical as it is associated with an infinite energy per mode, which renders it unnormalisable. However, it can be approached with arbitrary precision by means of simple families of Gaussian states.

A practical approximation of the EPR state is given by pure {\it two-mode squeezed states}, obtained by acting with the two-mode squeezing operator, Eq.~(\ref{tmsU}), on the vacuum state of a pair of modes $A$ and $B$. Using Eq.~(\ref{tmsS}), we find that a two-mode squeezed state with squeezing $r$ has vanishing first moments and covariance matrix
\begin{eqnarray}\label{tmsCM}
\sig_{AB}(r) &=& \gr{S}^{(2)}_{AB}(r) {\gr{S}^{(2)}_{AB}}\T(r) \\
&=& \nonumber \left(\begin{array}{cccc}
\cosh(2r)&0&\sinh(2r)&0\\
0&\cosh(2r)&0&-\sinh(2r)\\
\sinh(2r)&0&\cosh(2r)&0\\
0&-\sinh(2r)&0&\cosh(2r)
\end{array}\right)\! .
\end{eqnarray}
Notice that, from the Euler decomposition (\ref{euler1}), an alternative recipe to prepare two-mode squeezed states is to first apply local single-mode squeezing transformations to each mode (one squeezing in momentum, the other squeezing in position), and then let the two squeezed beams interfere at a balanced beam splitter. The result \be\gr{B}_{AB}(\mbox{$\frac{\pi}{4}$}) \big[\gr{S}_A^{(1)}(r,0) \oplus \gr{S}_B^{(1)}(r,\mbox{$\frac{\pi}{2}$})\big]\big[\gr{S}_A^{(1)}(r,0) \oplus \gr{S}_B^{(1)}(r,\mbox{$\frac{\pi}{2}$})\big]\T \gr{B}_{AB}\T(\mbox{$\frac{\pi}{4}$}) = \sig_{AB}(r)\ee is the same as Eq.~(\ref{tmsCM}). Different quantum optics laboratories across the world use either method to engineer such important states, which are also often referred to as `twin beams'.

To get a quantitative flavour of how well the two-mode squeezed states approximate the EPR state of \eq{eprstate}, we can define the EPR correlation parameter for a generic state $\rho$ as \begin{equation}\label{eprcorrelations}
\Upsilon(\rho) = \frac12 \left[\mathrm{Var}_\rho (\hat{q}_A-\hat{q}_B) + \mathrm{Var}_\rho (\hat{p}_A+\hat{p}_B)\right]\,.
\end{equation}
For the two-mode squeezed state of Eq.~(\ref{tmsCM}), the EPR correlations $\Upsilon$ amount to $e^{-2r}$, which tends to zero (the ideal EPR value) asymptotically for $r \rightarrow \infty$. At the time of writing the current manuscript (December 2013), the largest achievable two-mode squeezing in a stable optical configuration is about $10\ {\rm dB}$ \cite{stable10db}. In experimental papers, squeezing is often measured in deciBels, defined so that a squeezing degree $r$ corresponds to
\begin{equation}
\label{dB}
 10 \log_{10} [e^{2r}] \ {\rm dB}\,.
\end{equation}
Therefore, $10\ {\rm dB}$ corresponds to $r \approx 1.15$, i.e., to an EPR correlation parameter $\Upsilon \approx 0.1$.

The two-mode squeezed state is, quite naturally, the prototype of a CV entangled state, and is a central resource in many CV quantum information protocols, most significantly teleportation \cite{Telep,vaidman,Braunstein98,Furusawa98,vanlokfortshit,brareview}.

Considerable work in the literature has been spent and is currently being devoted to the characterisation of CV entanglement in general Gaussian states. Ref.~\cite{ourreview}, in particular, is a review article mostly focused on such a topic, being already quite didactic and suitable for a starter. In keeping with the spirit of this contribution, we are not going to reprise the majority of that material here (that would entail a considerable amount of cloning!). We will rather present some recent developments which make use of R\'enyi entropies to characterise Gaussian information and correlation measures.

We warn the reader that the remainder of this section is, therefore, part didactic and part reference.

Furthermore, when referring to correlation quantities for a density matrix $\rho_{AB}$ in this section, we shall adopt a notation which explicitly indicates the particular partition with respect to which correlations are computed. In particular, for total correlations and  entanglement, the subscript notation ``$A:B$'' is adopted to indicate that those correlations are shared between subsystems $A$ and $B$. For one-way classical and quantum correlations (discord), as defined later in the text, the directional notation ``$A|B$'' is adopted instead, to indicate ``$A$ {\it given} $B$'', i.e., to specify that we are looking at the change in the informational content of $A$ following a minimally disturbing marginal  measurement on $B$.

\subsection{Measures of information}

In quantum information theory, the degree of information contained in a quantum state $\rho$ is conventionally quantified via the von Neumann entropy of the state
\begin{equation}\label{vn}
{\cal S}(\rho) = - \text{Tr}(\rho \log \rho)\,,
\end{equation}
that is {\it the} direct counterpart to the Shannon entropy in classical information theory \cite{zykbook}. The most fundamental mathematical implications and physical insights in quantum information theory, ranging from the Holevo bound \cite{holevobound} all the way to the whole genealogy of quantum communication protocols \cite{merging,mother}, rely on a key property satisfied by the von Neumann entropy, the strong subadditivity inequality \cite{proofsss,hammer,Wehrl78}
\begin{equation}\label{eq:ssub}
{\cal S}(\rho_{AB}) + {\cal S}(\rho_{BC}) \geq {\cal S}(\rho_{ABC}) + {\cal S}(\rho_{B})\,,
\end{equation}
for an arbitrary tripartite state $\rho_{ABC}$.
The strong subadditivity inequality implies in particular that the mutual information
\begin{equation}\label{eq:mutual}
{\cal I}(\rho_{A:B}) = {\cal S}(\rho_A)+{\cal S}(\rho_B)-{\cal S}(\rho_{AB})\,,
\end{equation}
which is a measure of the total correlations between the subsystems $A$ and $B$ in the bipartite state $\rho_{AB}$,
is always nonnegative. Notice that, in Eq.~(\ref{eq:mutual}), ${\cal S}(\rho_{AB})$ denotes the global entropy of the state $\rho_{AB}$, while ${\cal S}(\rho_A)$ and ${\cal S}(\rho_B)$ correspond to the marginal entropies of the reduced states of subsystems $A$ and $B$ respectively.

{\it Entanglement} in a pure bipartite state is conventionally quantified by the von Neumann entropy of the reduced density matrix of one subsystem only, \be\label{veee}
{\cal E}(\ket{\psi}_{A:B}) = {\cal S}(\rho_A) = {\cal S}(\rho_B)\,,
 \ee
where $\rho_A = \text{Tr}_B[\ket{\psi}_{AB}\bra{\psi}]$ is the marginal state of $A$ obtained by partial tracing over $B$ (and vice versa for $\rho_B$).
 When evaluated on the two-mode squeezed state of Eq.~(\ref{tmsCM}), for instance \cite{ourreview}, the von Neumann entropy of entanglement $\cal E$ for $r \approx 1.15$ ($10 \ {\rm dB}$) is slightly bigger than $2$ ebits, i.e., the best current two-mode CV entangled states (in the optical Gaussian domain) have just a tad more entanglement than a pair of two-qubit Bell states: that's still a long way to the potentially infinite entanglement of the EPR state!

However, in classical as well as quantum information theory, several other entropic quantities have been introduced and studied. In particular, R\'{e}nyi-$\alpha$ entropies \cite{Renyi1970Book} are an interesting family of additive entropies, whose interpretation is related to derivatives of the free energy with respect to temperature \cite{baez}, and which have found applications especially in the study of channel capacities \cite{wolfeis,renyicapacity}, work value of information \cite{workvalue}, and entanglement spectra in many-body systems \cite{renyispectrum}. Entropies are in general measures of ignorance about the preparation of a state. R\'enyi-$\alpha$ entropies  are defined as
\begin{equation}\label{eq:ren}
{\cal S}_\alpha(\rho) = \frac{1}{1-\alpha} \log \big[\text{Tr} (\rho^\alpha)\big]\,,
\end{equation}
and reproduce  the von Neumann entropy in the limit $\alpha \rightarrow 1$.

\begin{figure}[t]
\centering
\includegraphics[width=10cm]{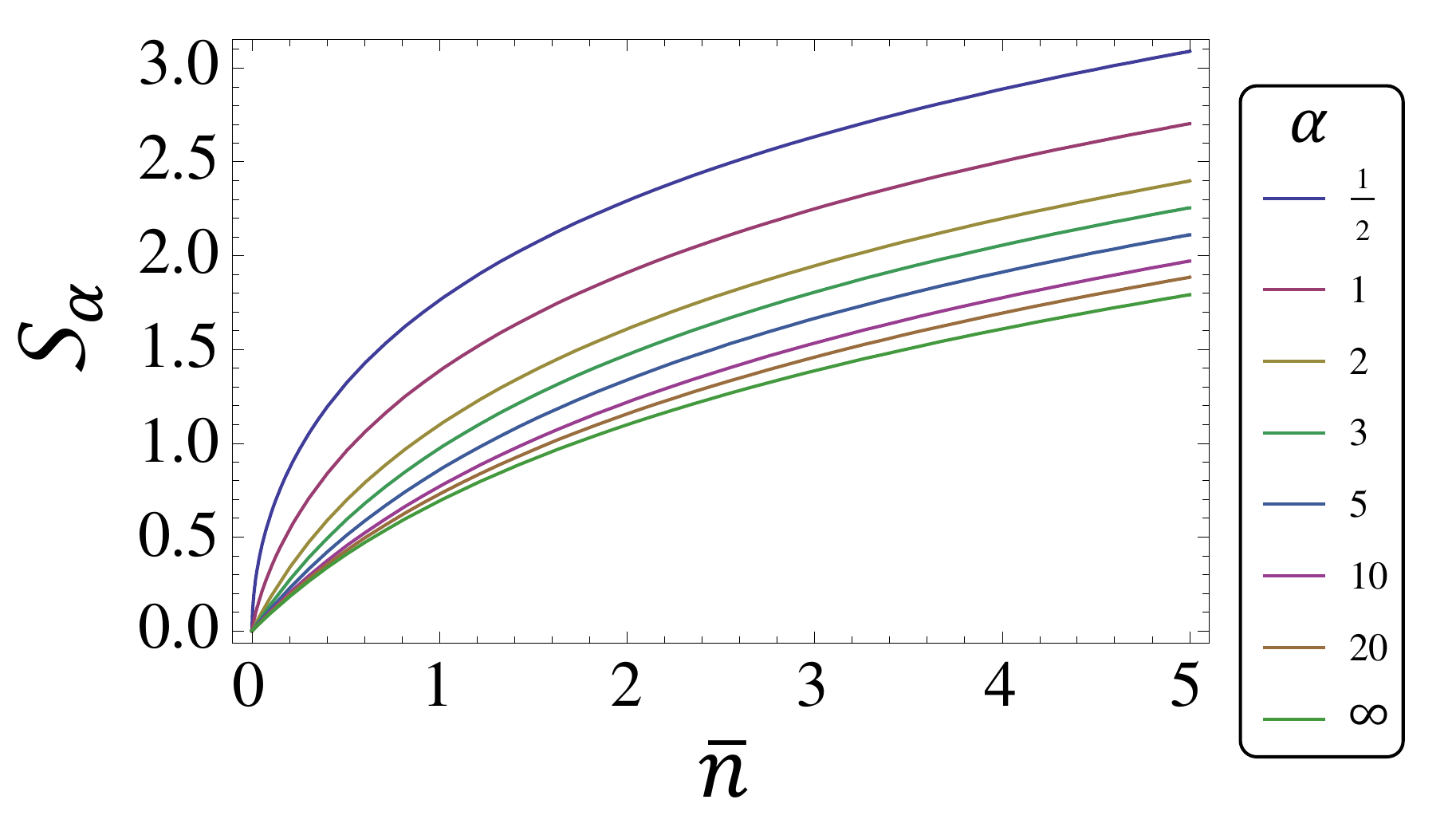}
\caption{R\'enyi-$\alpha$ entropies of a single-mode thermal state with mean particle number $\overline{n}$, plotted for $\alpha=\frac12,\,1,\,2,\,\ldots,\,\infty$ (from top to bottom).
\label{figren}}
\end{figure}

R\'enyi entropies can be evaluated on a generic $N$-mode Gaussian state $\rho$ in terms of its covariance matrix $\sig$ \cite{extremal}. We have (adopting from now on the natural base for the logarithm)
\begin{equation}\label{renyialg}
{\cal S}_\alpha(\rho) = \frac{\sum_{k=1}^N \ln[g_\alpha(\nu_k)]}{1-\alpha}\,,
\end{equation}
where $\{\nu_k\}$ are the symplectic eigenvalues of $\sig$, and \be g_\alpha(x)=2^{\alpha}/[(x+1)^\alpha-(x-1)^\alpha]\,.\ee The behaviour of the R\'enyi entropies for few values of $\alpha$ is plotted in Fig.~\ref{figren} for a single-mode thermal state (\ref{termique}) with covariance matrix $\sig=(2 \overline{n}+1) \gr{I}$ as a function of the mean particle number $\overline{n}$. All the R\'enyi entropies increase as expected with increasing temperature of the state (i.e., with increasing $\overline{n}$), while in general, for any given state, ${\cal S}_\alpha$ is a decreasing function of $\alpha$.

In Ref.~\cite{renyi} it has been demonstrated that a particular choice, $\alpha=2$, provides a natural and easily computable measure of information for any multimode Gaussian state.
The R\'enyi-$2$ entropy is directly related to the purity, and can be consequently computed very easily for a Gaussian state $\rho$,
\begin{equation}\label{eq:renyig}
{\cal S}_2(\rho) = - \ln \big[\text{Tr} (\rho^2) \big]= \frac12 \ln (\det \boldsymbol{\sigma})\,.
\end{equation}
This measure is operationally interpreted (modulo an additive constant) as the phase-space classical Shannon entropy ${H}$ of the Wigner distribution $W_{\rho}$ of the state $\rho$  (\ref{eq:wigG}), defined as \cite{Shannon48} \be {H}(W_{\rho}(\boldsymbol{\xi}))=-\int W_{\rho}(\boldsymbol{\xi}) \ln\{W_{\rho}(\boldsymbol{\xi})\} {\rm d}^{2N}\boldsymbol{\xi}.\ee Indeed, one has ${H}(W_{\rho}(\boldsymbol{\xi}))={\cal S}_2(\rho)+N(1+\ln \pi)$ \cite{renyi}.
A crucial property of ${\cal S}_2$ in the Gaussian scenario (and only there, as this is not true for other states, not even qubit states!) is that it fulfills a condition analogous to the one in (\ref{eq:ssub}) for the von Neumann entropy.
Let $\rho_{ABC}$ be a tripartite Gaussian state whose subsystems encompass arbitrary number of modes. Writing its covariance matrix in block form as in Eq.~(\ref{CM}), and using the definition (\ref{eq:renyig}), we have the following \cite{renyi}
\begin{theorem}\label{teo}
The R\'{e}nyi-$2$ entropy ${\cal S}_2$ satisfies the strong subadditivity inequality for all Gaussian states $\rho_{ABC}$,
\begin{equation}\label{eq:ss2}\begin{split}
 {\cal S}_2(\rho_{AB}) + {\cal S}_2(\rho_{BC}) - {\cal S}_2(\rho_{ABC}) - {\cal S}_2(\rho_{B}) \\
\quad= \frac12 \ln \left( \frac{\det\sig_{AB} \det\sig_{BC}}{\det\sig_{ABC}\det\sig_B} \right) \geq 0\,.
\end{split}\end{equation}
\end{theorem}
\noindent {\it Proof.} The result follows by applying a particular norm compression inequality to the covariance matrix $\sig_{ABC}$. Given a positive  Hermitian matrix $\gr A \in \mathbb{M}_m$, and given any two index sets $\alpha, \beta \subseteq N=\{1,\ldots,m\}$, the Hadamard-Fisher inequality \cite{hornjohnson} states that
$\det \gr A_{\alpha \cup \beta} \det \gr A_{\alpha \cap \beta} \leq \det \gr A_\alpha \det \gr A_\beta$.
Recalling that any covariance matrix $\sig_{ABC}$ is a positive  real symmetric matrix \cite{williamson}, the claim follows upon identifying $\alpha$ with the indices of modes $AB$ and $\beta$ with the indices of modes $BC$. \hfill $\Box$

Strong subadditivity  ``is a potent hammer in the quantum information theorist's toolkit'' \cite{hammer}.
Beyond its apparent simplicity, Theorem \ref{teo} has profound consequences. It yields that the  core of quantum information theory can be consistently reformulated, within the Gaussian world \cite{pirandolareview}, using the simpler and physically natural  R\'{e}nyi-$2$ entropy in alternative to the von Neumann one.

\subsection{R\'enyi-$2$ measures of correlations}
 In the rest of this section we shall focus on defining Gaussian R\'{e}nyi-$2$ quantifiers of entanglement and other correlations for Gaussian states \cite{renyi}.
\subsubsection{Total correlations} For a bipartite Gaussian state $\rho_{AB}$ with covariance matrix as in \eq{eq:cms}, the  total correlations between subsystems $A$ and $B$ can be quantified by the R\'enyi-2 mutual information ${\cal I}_2$,
defined as \cite{renyi}
\begin{eqnarray}\label{eq:remutual}
{\cal I}_2(\rho_{A:B}) &=& {\cal S}_2(\rho_A) + {\cal S}_2(\rho_B) - {\cal S}_2(\rho_{AB}) \nonumber \\
&=&\frac12 \ln\left(\frac{\det\boldsymbol{\sigma}_A \det\boldsymbol{\sigma}_B}{\det\boldsymbol{\sigma}_{AB}}\right)\,,
\end{eqnarray}
which measures the phase space distinguishability between the Wigner function of $\rho_{AB}$ and the Wigner function associated to the product of the marginals $\rho_A \otimes \rho_B$ (which is, by definition, a state in which the subsystems $A$ and $B$ are completely uncorrelated). Notice that, from Theorem~\ref{teo}, the quantity ${\cal I}_2(\rho_{A:B})$ is always nonnegative, and vanishes if and only if $\rho_{AB}=\rho_A \otimes \rho_B$, i.e., $\sig_{AB} = \sig_A \oplus \sig_B$.

\subsubsection{Entanglement}
In the previous paragraph, we spoke about total correlations, but that is not the end of the story. In general, we can discriminate between classical and quantum correlations \cite{entanglement,modireview}.

A bipartite pure state $|\psi_{AB}\rangle$ is quantum-correlated, i.e., is `entangled', if and only if it cannot be factorised as $|\psi_{AB}\rangle =|\phi_A\rangle\otimes |\chi_B\rangle$.   On the other hand, a mixed state $\rho_{AB}$ is entangled if and only if it cannot be written as \begin{equation}\label{sep}
\rho_{AB}=\sum_i p_i \varrho_{A_i}\otimes \varrho_{B_i},\end{equation} that is a convex combinations of product states, where $\{p_i\}$ are probabilities and $\sum_i p_i=1$. Unentangled states are called `separable'. The reader can refer to Ref.~\cite{entanglement} for an extensive review on entanglement. In particular, one can quantify the amount of entanglement in a state by building specific measures. For Gaussian states, any measure of entanglement will be a function of the elements of the covariance matrix only, since the displacement vector can be nullified by local unitaries alone.

There are a number of meaningful entanglement measures for Gaussian states. Conventionally, two families of measures have been studied; one family encompasses so-called {\it negativities} (including prominently the logarithmic negativity \cite{vidwer02}). These measures are computable for any multimode Gaussian state $\rho$ and quantify the violation of a particular separability criterion based on the positivity of the partial transposition \cite{peres96,horodecki96,Simon00} of $\rho$. An extensive and instructive account of negativities for Gaussian states is available in \cite{ourreview} and will not be included here.

The second way to quantify Gaussian entanglement is to construct so-called (Gaussian) convex roof extended measures.
Given an entanglement monotone $E$ for pure states $\ket{\psi}$, it can be extended to mixed states  $\rho$ by means of the convex roof construction,
\begin{equation}\label{croof}
E(\rho) = \inf_{\{p_i,\ket{\psi_i}\}} \sum_i p_i E(
(\ket{\psi_i}) \, ,
\end{equation}
where the minimisation is taken over all the decompositions of $\rho$ into ensembles of pure states $\{\ket{\psi_i}\}$ with  probabilities $\{p_i\}$,
$\rho = \sum_i p_i \ketbra{\psi_i}{\psi_i}$. If $E$ is chosen, for instance, as the conventional von Neumann entropy of entanglement (\ref{veee}), then the measure in Eq.~(\ref{croof}) defines the {\it entanglement of formation} \cite{Bennett96pra,entanglement}. In general, convex roof extended measures quantify how much entanglement, on average, needs to be expended to create $\rho$ out of an ensemble of pure states, choosing the cheapest realisation. Notice that for CV systems, the sum can be replaced by an integral and the discrete $\{p_i\}$'s by a probability distribution. For Gaussian states, one can define upper bounds to the absolute minimum in (\ref{croof}), by restricting the minimisation to a smaller set, namely to decompositions of a Gaussian $\rho$ into pure {\it Gaussian} states $\ket{\psi_i}$ only. For any mixed Gaussian $\rho$, there exists at least one such decomposition, which is realised in terms of pure coherent states, with displacement distributed according to a Gaussian distribution in phase space, see e.g.~\cite{geof}. The problem of identifying the optimal decomposition in the Gaussian convex roof, for a given entanglement monotone $E$ and an arbitrary Gaussian state $\rho$ with covariance matrix $\sig$, can be reformulated quite nicely in terms of $\sig$ alone \cite{geof,ordering,ourreview}.
The entanglement measure we consider here belongs to this second class, and is determined by the use, as pure-state monotone $E$, of the R\'enyi-$2$ entropy of entanglement.

Precisely, a measure of bipartite entanglement ${\cal E}_2$  for Gaussian states based on R\'enyi-$2$ entropy can be defined as follows \cite{renyi}. Given a  Gaussian state $\rho_{AB}$ with covariance matrix $\sig_{AB}$, we have
\begin{equation}\label{eq:GR2_ent}
{\cal E}_2 (\rho_{A:B}) = \inf_{\left\{\gam_{AB}\ |\ 0<\gam_{AB} \le \sig_{AB}, \, \det{\gam_{AB}}=1\right\}} \frac12 \ln \left(\det \gam_A\right)\,,
\end{equation}
where the minimisation is over pure $N$-mode Gaussian states with covariance matrix $\gam_{AB}$ smaller than $\sig_{AB}$\footnote{For two real symmetric matrices $\gr{M}$ and $\gr{N}$, the statement ${\gr M} \leq {\gr N}$ means that $\gr{N}-\gr{M} \geq 0$, i.e., that the matrix $\gr{N}-\gr{M}$ has all nonnegative eigenvalues.}, and $\gam_A$ refers to the marginal covariance matrix of subsystem $A$ obtained from $\gam_{AB}$ by partial tracing over the modes of subsystem $B$.
For a pure Gaussian state $\rho_{AB} = \ket{\psi_{AB}}\bra{\psi_{AB}}$ with covariance matrix $\sig_{AB}^{\rm pure}$, the minimum is saturated by $\gam_{AB}=\sig_{AB}^{\rm pure}$, so that the measure of \eq{eq:GR2_ent} reduces to the pure-state R\'enyi-$2$ entropy of entanglement,
\begin{equation}\label{eq:GR2_ent_pure}
{\cal E}_2(\sig_{A:B}^{\rm pure})= {\cal S}_2(\sig_A) = \frac12 \ln (\det{\sig_A})\,,
\end{equation}
where $\sig_A$ is the reduced covariance matrix of subsystem $A$.
For a generally mixed state, Eq.~(\ref{eq:GR2_ent})  amounts to taking the Gaussian convex roof of the pure-state R\'{e}nyi-$2$ entropy of entanglement, according to the formalism of \cite{geof}. Closed formulae for ${\cal E}_2$ can be obtained for special classes of two-mode Gaussian states \cite{renyi}, and will be reported later in this section.

In principle, one might expect that convex decompositions of a generic mixed Gaussian state over ensembles of pure non-Gaussian states might lead to a further minimisation of the average entanglement, i.e., that the Gaussian convex roof might only be a non-tight upper bound to the true value of the entanglement measures for mixed Gaussian states. A longstanding conjecture in CV quantum information theory, related to the so-called bosonic additivity conjecture \cite{Holevo01,pirandolareview}, has postulated that this is not the case, i.e., that Gaussian decompositions {\it are} optimal for entanglement measures of Gaussian states based on convex roof constructions (\ref{croof}) using the von Neumann entropy (corresponding to the canonical entanglement of formation \cite{entanglement,Bennett96pra,giedke03,geof}) as well as all the other R\'enyi entropies. Remarkably, during completion of the present article, this key conjecture has been finally proven true by Giovannetti and coworkers \cite{giovaproof1,giovaproof2,giovaproof3}.  Among a number of important implications for practical quantum communication, this entails that the R\'enyi-$2$ entanglement measure defined above is an additive entanglement monotone.

\subsubsection{Classical correlations}
For pure states, entanglement is the only kind of quantum correlation. A pure separable state is essentially classical, and the subsystems display no correlation at all. On the other hand, for mixed states, one can identify a finer distinction between classical and quantum correlations, such that even most separable states display a definite quantum character \cite{zurek,vedral}.

Conceptually, one-way classical correlations are those extractable by local measurements; they can be defined in terms of how much the ignorance about the state of a subsystem, say $A$, is reduced when the most informative local measurement is performed on subsystem $B$ \cite{vedral}. The quantum correlations (known as {\it quantum discord}) are, complementarily, those destroyed by local measurement processes, and correspond to the change in total correlations between the two subsystems, following the action of a minimally disturbing local measurement on one subsystem only \cite{zurek}. For Gaussian states, R\'enyi-$2$ entropy can be adopted once more to measure ignorance and correlations \cite{renyi}.

To begin with, we can introduce a Gaussian R\'enyi-$2$ measure of one-way classical correlations \cite{vedral,adessodatta,giordaparis,renyi}. We define  ${\cal J}_2(\rho_{A|B})$ as the maximum decrease in the R\'{e}nyi-$2$ entropy of subsystem $A$, given a Gaussian measurement has been performed on subsystem $B$, where the maximisation is over all Gaussian measurements [see Eqs.~\ref{gpovm},(\ref{eq:schur})].
We have then
\begin{eqnarray}\label{eq:J2}
{\cal J}_2(\rho_{A|B}) &=& \sup_{\gr\Gamma_B^\Pi} \frac12 \ln \left(\frac{\det \sig_A}{\det\tilde{\gr\sigma}^{\Pi}_A}\right)\,; \nonumber \\ & & \\
{\cal J}_2(\rho_{B|A}) &=& \sup_{\gr\Gamma_A^\Pi} \frac12 \ln \left(\frac{\det \sig_B}{\det\tilde{\gr\sigma}^{\Pi}_B}\right)\,, \nonumber
\end{eqnarray}
where the one-way classical correlations ${\cal J}_2(\rho_{B|A})$, with Gaussian measurements on $A$, have been defined accordingly by swapping the roles of the two subsystems, $A \leftrightarrow B$. Notice that, for the same state $\rho_{AB}$,  ${\cal J}_2(\rho_{A|B}) \neq {\cal J}_2(\rho_{B|A})$ in general: The classical correlations depend on which subsystem is measured.

For two-mode Gaussian states, it can be proven that the classical correlations always exceed entanglement, $\{{\cal J}_2(\rho_{A|B}),{\cal J}_2(\rho_{B|A})\} \geq {\cal E}_2(\rho_{A:B})$ \cite{renyi}.

\subsubsection{Quantum correlations (Discord)}
We can now define a Gaussian measure of quantumness of correlations based on R\'{e}nyi-$2$ entropy.
Following the landmark study by Ollivier and Zurek \cite{zurek}, and the recent investigations of Gaussian quantum discord \cite{adessodatta,giordaparis,renyi}, we define the R\'enyi-$2$ discord as the difference between mutual information (\ref{eq:remutual}) and classical correlations (\ref{eq:J2}),
\begin{eqnarray}\label{eq:D2}
{\cal D}_2(\sig_{A|B}) &=& {\cal I}_2(\sig_{A:B})- {\cal J}_2(\sig_{A|B}) \nonumber\\
&=&\inf_{\gr\Gamma_B^\Pi} \frac12 \ln \left(\frac{\det \sig_B \det \tilde{\gr\sigma}^{\Pi}_A}{\det \sig_{AB}}\right)\,;
\nonumber \\ & & \\
{\cal D}_2(\sig_{B|A}) &=& {\cal I}_2(\sig_{A:B})- {\cal J}_2(\sig_{B|A}) \nonumber \\
&=&\inf_{\gr\Gamma_A^\Pi} \frac12 \ln \left(\frac{\det \sig_A \det \tilde{\gr\sigma}^{\Pi}_B}{\det \sig_{AB}}\right)\,.
\nonumber
\end{eqnarray}
The discord is clearly a nonsymmetric quantity as well. It captures general quantum correlations even in the absence of entanglement \cite{zurek,modireview}.
An interesting fact is that all Gaussian states possess a nonzero discord, apart from product states which are completely uncorrelated \cite{adessodatta,ralphdiscord}.

\subsubsection{Remarks}

Let us remark that we have defined classical and quantum correlations by restricting the optimisation over Gaussian measurements only. This means that, potentially allowing for more general non-Gaussian measurements (e.g., photon counting), it would seem one could possibly obtain higher classical correlations and lower quantum ones.
Until recently, there was numerical and partial analytical evidence to support the conclusion that for two-mode Gaussian states, Gaussian measurements are optimal for the calculation of general one-way classical and quantum discord \cite{adessodatta,allegramente}. With the aforementioned long-sought proof of the bosonic additivity conjecture, only very recently reported \cite{giovaproof1,giovaproof2,giovaproof3}, this matter has been settled:  No non-Gaussian measurements can further reduce the value of the quantum discord for Gaussian states (see also \cite{pirlaimplications}). Notably, one can therefore obtain \textit{optimal} closed analytical expressions for Eqs.~(\ref{eq:J2}) and (\ref{eq:D2}) for the case of $A$ and $B$ being single modes, that is, $\rho_{AB}$ being a general two-mode Gaussian state \cite{adessodatta,renyi}, as reported later in explicit form.

Finally, let us observe that \begin{eqnarray}\label{pureeq}
\frac{{\cal I}_2(\rho^{\rm pure}_{A:B})}2&=&{\cal J}_2(\rho^{\rm pure}_{A|B})={\cal J}_2(\rho^{\rm pure}_{B|A})={\cal D}_2(\rho^{\rm pure}_{A|B})={\cal D}_2(\rho^{\rm pure}_{B|A})\nonumber \\ &=&{\cal E}_2(\rho^{\rm pure}_{A:B})={\cal S}_2(\rho_A)={\cal S}_2(\rho_B)\,,
\end{eqnarray}
for {\it pure} bipartite Gaussian states $\rho_{AB}^{\rm pure}=\ket{\psi}_{AB}\bra{\psi}$ of an arbitrary number of modes. That is, general quantum correlations reduce to entanglement, and an equal amount of classical correlations is contained as well in pure states.

\subsubsection{Explicit expressions for two-mode states}

\paragraph{Standard form.}
The covariance matrix $\sig_{AB}$ of any two-mode Gaussian state $\rho_{AB}$ can be transformed, by means of local unitary (symplectic) operations, into a standard form of the type \cite{Duan00,ourreview}
\begin{equation}
\label{eq:gamsf}
\sig_{AB}=\left(\begin{array}{cc}
{\sig_A}&{\gr\varepsilon_{AB}}\\
{\gr\varepsilon}_{AB}^{\sf T}&{\sig_B}
\end{array}\right) = \left(\begin{array}{cccc}
a&0&c_{+}&0\\
0&a&0&c_{-}\\
c_{+}&0&b&0\\
0&c_{-}&0&b
\end{array}\right)\;,
\end{equation}
where $a,b\geq 1$, $\left[\left(a^2-1\right) \left(b^2-1\right)-2 c_- c_+-a b c_+^2+c_-^2 \left(-a b+c_+^2\right)\right] \geq 0$, and we can set $c_+ \ge |c_-|$ without losing any generality. These conditions ensure that the {\it bona fide} condition (\ref{bonfide}) is verified. For pure Gaussian states, $b=a$, $c_+=-c_-=\sqrt{a^2-1}$, i.e., any pure two-mode Gaussian state is equivalent (up to local unitaries) to a two-mode squeezed state of the form (\ref{tmsCM}).

All the formulae presented in the following will be written explicitly for standard form covariance matrices for simplicity. However, they can be recast in a locally invariant form by expressing them in terms of the four local symplectic invariants of a generic two-mode Gaussian state \cite{Serafini2004JPB}, \begin{equation}
\begin{split}
I_1&=\det \sig_A,\\
I_2&=\det \sig_B,\\
I_3&=\det{\gr\varepsilon}_{AB},\\
I_4&=\det\sig_{AB}.
 \end{split}\end{equation}
 This is accomplished by inverting the relations \be I_1=a^2,\ \  I_2=b^2,\ \  I_3=c_+ c_-,\ \  I_4=(a b -c_+)(a b - c_-), \ee so that the invariants $\{I_j\}_{j=1}^4$ appear explicitly in the formulae below \cite{ourreview}. The obtained expressions would then be valid for two-mode covariance matrices in any symplectic basis, beyond the standard form.

\paragraph{R\'enyi-$2$ entanglement.} For generally mixed two-mode Gaussian states $\rho_{AB}$, the R\'{e}nyi-$2$ entanglement measure ${\cal E}_2(\rho_{A:B})$, defined by Eq.~(\ref{eq:GR2_ent}), admits the following expression if the covariance matrix $\sig_{AB}$ is in standard form
\cite{geof,ordering},
\begin{equation}\label{eq:E2AB}
{\cal E}_2(\rho_{A:B}) = \frac12 \ln \left(\inf_{\theta \in [0,2\pi]}  m_\theta (a,b,c_+, c_-)\right)\,,
\end{equation}
with
\begin{eqnarray}\label{eq:mfunc}
&&\!\!\!\!m_\theta (a,b,c_+, c_-)\ =\ 1 +
\scriptstyle{\left[c_+(ab-c_-^2)-c_-+\cos \theta \sqrt{\left[a -
b(ab-c_-^2)\right]\left[b-a(ab-c_-^2)\right]}\right]^2}
\nonumber \\
&&\times\scriptstyle{\left\{
2\left(ab-c_-^2\right)\left(a^2+b^2+2c_+c_- \right) +\ \sin \theta\left(a^2-
b^2\right)\sqrt{1-\frac{\left[c_+(ab-c_-^2)+c_-\right]^2}{\left[a -
b(ab-c_-^2)\right]\left[b-a(ab-c_-^2)\right]}}\right.} \nonumber \\
&&\scriptstyle{\left.\ -\ \frac{\cos \theta\left[2abc_-^3+\left(a^2+
b^2\right)c_+c_-^2+\left(\left(1-2b^2\right)a^2+
b^2\right)c_--ab\left(a^2+b^2- 2\right)c_+\right]}{\sqrt{\left[a -
b(ab-c_-^2)\right]\left[b-a(ab-c_-^2)\right]}} \right\}^{-1}}.
\end{eqnarray}
The optimal $\theta$ minimising Eq.~(\ref{eq:mfunc}) can be found numerically for general two-mode Gaussian states \cite{geof}, and analytically for relevant subclasses of states (including symmetric states \cite{giedke03}, squeezed thermal states, and states with one symplectic eigenvalue equal to $1$ \cite{ordering}).

\paragraph{R\'enyi-$2$ classical correlations and discord.}
For generally mixed two-mode Gaussian states $\rho_{AB}$, the R\'{e}nyi-$2$ measures of one-way classical correlations ${\cal J}_2(\rho_{A|B})$ and quantum discord ${\cal D}_2(\rho_{A|B})$, defined by Eqs.~(\ref{eq:J2}) and (\ref{eq:D2}), respectively, admit the following expression if the covariance matrix $\sig_{AB}$ is in standard form \cite{adessodatta}
\begin{eqnarray}
{\cal J}_2(\rho_{A|B}) &=& \ln a -    \frac12 \ln \left(\inf_{\lambda,\varphi}{\det\tilde{\gr\sigma}^{\Pi_{\lambda,\varphi}}_A}\right)\,, \label{eq:J2AB} \\
{\cal D}_2(\rho_{A|B}) &=& \ln b - \frac12 \ln \big(\det{\sig_{AB}}\big) +   \frac12 \ln \left(\inf_{\lambda,\varphi}{\det\tilde{\gr\sigma}^{\Pi_{\lambda,\varphi}}_A}\right)\,, \label{eq:D2AB}
\end{eqnarray}
with $\lambda \in (0,\infty),\,\varphi\in[0,2\pi]$, and
\begin{equation}\label{eq:detcond}
\det\tilde{\gr\sigma}^{\Pi_{\lambda,\varphi}}_A =  \scriptscriptstyle{\frac{2 a^2 (b + \lambda) (1 + b \lambda)-a \left(c_+^2+c_-^2\right) \left(2 b \lambda +\lambda ^2+1\right) +2 c_+^2 c_-^2 \lambda +a \left(c_+^2-c_-^2\right) \left(\lambda ^2-1\right) \cos (2 \varphi)}{{2 (b + \lambda) (1 + b \lambda)}}}\,.
\end{equation}
The optimal values of $\lambda$ and $\varphi$ minimising Eq.~(\ref{eq:detcond}) can be found analytically for all two-mode Gaussian states \cite{adessodatta}. In particular, for standard form covariance matrices, one gets
\begin{eqnarray}\label{eq:optdetcond}
\!\!\!&&\inf_{\lambda,\varphi}\det\tilde{\gr\gamma}^{\Pi_{\lambda,\varphi}}_A = \\
&&\quad \ \ \left\{
\begin{array}{l}
 a \left(a-\frac{c_+^2}{b}\right)\,, \\ \qquad \qquad \qquad \qquad  \text{if}\quad {\scriptstyle{\left(a b^2 c_-^2-c_+^2 \left(a+b c_-^2\right)\right) \left(a b^2 c_+^2-c_-^2 \left(a+b c_+^2\right)\right)<0}}\, \mbox{;} \\ \quad \\
 \scriptscriptstyle{\frac{2 \left|c_- c_+\right| \sqrt{\left(a \left(b^2-1\right)-b c_-^2\right) \left(a \left(b^2-1\right)-b c_+^2\right)}+\left(a \left(b^2-1\right)-b c_-^2\right) \left(a \left(b^2-1\right)-b c_+^2\right)+c_-^2 c_+^2}{\left(b^2-1\right)^2}} \,, \\
 \qquad \qquad \qquad \qquad  \text{otherwise.}
\end{array}
\right. \nonumber
\end{eqnarray}
Inserting Eq.~(\ref{eq:optdetcond}) into Eqs.~(\ref{eq:J2AB},\ref{eq:D2AB}) one gets closed formulae for the one-way  classical correlations and for the  discord of general two-mode Gaussian states based on R\'enyi-$2$ entropy.

\subsubsection{Tripartite entanglement}
Multipartite entanglement can be defined as well for Gaussian states in terms of R\'enyi-$2$ entropy. In particular, a measure of genuine tripartite entanglement can be defined from the property that entanglement is `monogamous' \cite{ckw}. A didactic excursus on the monogamy of entanglement can be found, e.g., in \cite{pisa}.

For an entanglement monotone $E$ and a $n$-partite state $\rho_{A_1 A_2 \ldots A_n}$, the monogamy relation (choosing party $A_1$ as the focus), which constrains the distribution of bipartite entanglement among different splits, can be written as \cite{ckw} \be E(\rho_{A_1:A_2 \ldots A_n}) - \sum_{j=2}^{n} E(\rho_{A_1:A_j}) \geq 0\,.\ee
Counterintuitively, not all entanglement monotones obey the monogamy inequality as just formalised.
Entanglement measures based on  R\'{e}nyi-$\alpha$ entropies (for $\alpha \geq 2$) \cite{barrylagallina}, as well as the tangle (squared concurrence) \cite{ckw,osborne}, satisfy this inequality for general $n$-qubit states. A Gaussian version of the tangle (based on squared negativity \cite{ourreview}) has been defined that obeys the inequality for all $N$-mode Gaussian states \cite{hiroshimaa}. It can be shown that ${\cal E}_2$ does too \cite{renyi}: the R\'enyi-$2$ entanglement defined in Eq.~(\ref{eq:GR2_ent}) is monogamous for all $n$-mode Gaussian states $\rho_{A_1 A_2 \ldots A_n}$,
\begin{equation}\label{eq:mono}
\begin{array}{c}{\cal E}_2(\rho_{A_1:A_2 \ldots A_n}) - \sum_{j=2}^{n} {\cal E}_2(\rho_{A_1:A_j}) \geq 0\,,\end{array}
\end{equation}
where each $A_j$ comprises one mode only.

 Referring the interested reader to \cite{ourreview,contangle,3mpra,hiroshimaa,renyi} for further details, we focus here on the consequences of this inequality for  pure three-mode Gaussian states.
 Up to local unitaries, the covariance matrix $\sig_{A_1A_2A_3}$ of any pure three-mode Gaussian state can be written in the following standard form \cite{3mpra}
 \begin{eqnarray}\label{CM3}
 \sig_{A_1A_2A_3}=\left( \begin{array}{cccccc}
 a_1 & 0 &c_3^+&0&c_2^+&0\\
0 & a_1 & 0&c_3^-&0&c_2^-\\
c_3^+& 0 &a_2&0&c_1^+&0\\
0 & c_3^- & 0&a_2&0&c_1^-\\
c_2^+ & 0 &c_1^+&0&a_3&0\\
0 & c_2^-& 0&c_1^-&0&a_3\\
 \end{array}
   \right)
\end{eqnarray}
where
\begin{eqnarray}
&&\!\!\!\!\!\!\!\!\!\! 4\sqrt{a_j a_k}c_i^\pm =  \sqrt{[(a_i-1)^2-(a_j-a_k)^2][(a_i+1)^2-(a_j-a_k)^2]} \nonumber \\
&&\!\!\!\!\!\!\!\!\!\!\qquad \qquad \ \  \pm\sqrt{[(a_i-1)^2-(a_j+a_k)^2][(a_i+1)^2-(a_j+a_k)^2]}\,, \\
&&\!\!\!\!\!\!\!\!\!\! \hbox{and } |a_j - a_k|+1\leq a_i\leq a_j+a_k-1\,, \nonumber
\end{eqnarray}
with $\{i,j,k\}$ being all possible permutations of $\{1,2,3\}$.

The R\'enyi-$2$ entanglement in the two-mode reduced state with covariance matrix $\sig_{A_i A_j}$ is
\begin{equation}
{\cal E}_2(\rho_{A_i:A_j}) = \frac12 \ln g_k \,,
\end{equation}
with \cite{ordering}
\begin{equation}\label{eq:mglems}
g_k = \left\{
                  \begin{array}{ll}
                    1, & \hbox{if $a_k \geq \sqrt{a_i^2+a_j^2-1}$;} \\
                    \displaystyle\frac{\beta}{8 a_k^2}, & \hbox{if $\alpha_k < a_k < \sqrt{a_i^2+a_j^2-1}$;} \\
                    \displaystyle\left(\frac{a_i^2-a_j^2}{a_k^2-1}\right)^2, & \hbox{if $a_k \leq \alpha_k$.}
                  \end{array}
                \right.
\end{equation}
Here we have set
\begin{eqnarray}
\alpha_k&=&{\small{\sqrt{\frac{2(a_i^2+a_j^2)+(a_i^2-a_j^2)^2+|a_i^2-a_j^2|\sqrt{(a_i^2-a_j^2)^2+8(a_i^2+a_j^2)}}{2(a_i^2+a_j^2)}}}}\,,\nonumber\\
\beta&=&{\small{2 a_1^2+2 a_2^2+2 a_3^2+2 a_1^2 a_2^2+2 a_1^2 a_3^2+2 a_2^2 a_3^2-a_1^4-a_2^4-a_3^4-\sqrt{\delta}-1}}\,,\nonumber\\
\delta&=&{\small{\mbox{$\prod_{\mu,\nu=0}^{1}$} \big[(a_1+(-1)^\mu a_2 + (-1)^\nu a_3)^2-1\big]}}\,.
\end{eqnarray}

\begin{figure}[t]
\centering
\includegraphics[width=9.5cm]{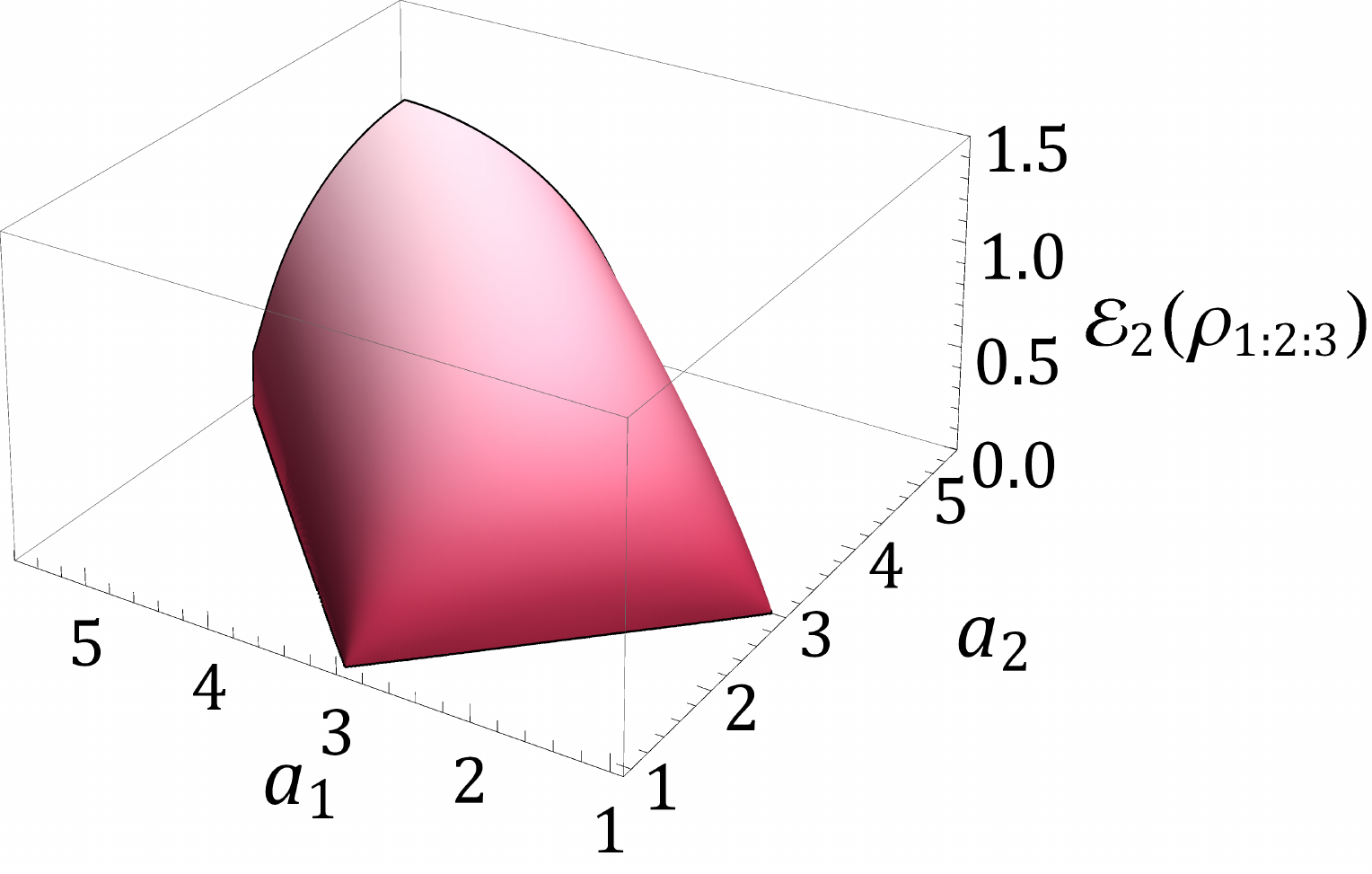}
\caption{Genuine tripartite R\'enyi-2 entanglement ${\cal E}_2(\rho_{1:2:3})$, Eq.~(\ref{renyi3}), of three-mode pure Gaussian states with covariance matrix (\ref{CM3}) plotted versus the single-mode parameters $a_1$ and $a_2$ at fixed $a_3=3$.
\label{fig3}}
\end{figure}

We can define the {\it residual tripartite} R\'enyi-$2$ entanglement, with respect to the focus mode $A_i$, as
\begin{eqnarray}\label{renyi3}
{\cal E}_2(\rho_{A_i:A_j:A_k})&=& {\cal E}_2(\rho_{A_i:A_jA_k}) - {\cal E}_2(\rho_{A_i:A_j}) - {\cal E}_2(\rho_{A_i:A_k}) \nonumber \\
&=& \frac12 \ln \left(\frac{a_i^2}{g_k\ g_j}\right)\,.
\end{eqnarray}
In general, this expression (see Fig.~\ref{fig3}) is dependent on the choice of the focus mode. Nevertheless, let us consider the particular case of a fully inseparable three-mode pure Gaussian state such that entanglement is nonzero for all global mode splittings and for all reduced two-mode bipartitions, ${\cal E}_2(\rho_{A_i:A_jA_k})>0$, ${\cal E}_2(\rho_{A_i:A_j})>0$, $\forall \{i,j,k\}$. In our parametrisation, this occurs when \cite{ordering} \begin{equation}
|a_i-a_j|+1<a_k < \sqrt{a_i^2+a_j^2-1}\,,\end{equation} for all mode permutations. It is immediate to see that the simultaneous verification of such a condition for all mode permutations imposes $a_k > \alpha_k$, $\forall k=1,2,3$. In this case, exploiting Eq.~(\ref{eq:mglems}), the residual tripartite R\'enyi-$2$ entanglement becomes
\begin{equation}
{\cal E}_2(\rho_{A_i:A_j:A_k}) = \frac12 \ln \left( \frac{64 a_i^2 a_j^2 a_k^2}{\beta^2} \right)\,,
\end{equation}
which is manifestly invariant under mode permutations. This is analogous to the case of the three-tangle for pure three-qubit states \cite{ckw}, which is an invariant function.
The  symmetry in the Gaussian tripartite entanglement is however broken on states for which some of the reduced two-mode bipartitions become separable.
A comparison between genuine tripartite R\'enyi-$2$ entanglement and genuine tripartite Bell nonlocality for three-mode Gaussian states has been recently performed \cite{samy}.

For $N>3$ modes, measures of {\it genuine} Gaussian $N$-partite entanglement may be defined from a refinement of the conventional monogamy inequality (\ref{eq:mono}), which includes a decomposition of the residual non-pairwise entanglement into independent $K$-partite contributions involving groups of $K=3,\ldots,N$ modes. Such a {\it strong} monogamy inequality (which so far has seen no analogue for qubit systems) has been established for permutationally invariant $N$-mode Gaussian states using a different entanglement measure in \cite{strongmono}, but we believe it can be proven as well (even leading to somehow handier formulas for the genuine $N$-partite entanglement) by using the R\'enyi-$2$ entanglement. While we have not attempted such a proof for lack of time, we invite a  reader of good will to face the task rigorously (and report back to us!).

 \section{Conclusions and outlook}\label{secBeyond}
 Gaussian, Gaussian everywhere. For the last one and a half decades, the vast majority of CV quantum information has revolved around Gaussian states (in theory and in experiments), including most of our own research. Gaussian states and Gaussian operations represent however a tiny null-measure corner in the infinite jungle of CV systems, and we are now aware of a number of tasks where they simply are not enough.

 For starters, it is impossible to distill Gaussian entanglement by Gaussian operations \cite{nogo1,nogo2,nogo3}. Other no-go results are known for bit commitment \cite{cerfbit} and error correction \cite{nogoerr}. To estimate parameters pertaining to Gaussian evolutions in quantum metrology, Gaussian probes are typically not optimal \cite{mypraest}. More generally, Gaussian states have a fundamental limitation which follows from their {\it extremality}: they are the least entangled states (according to suitable entanglement monotones) among all states of CV systems with given first and second
moments \cite{extra}. Experimentally, it has been recently demonstrated \cite{nongaussexp,furunphot2010}
that a two-mode squeezed Gaussian state can be ``degaussified'' by
coherent subtraction of single photons, resulting in a mixed
non-Gaussian state whose nonlocal properties and entanglement degree
are enhanced. Suitable non-Gaussian states can lead to higher teleportation fidelities for classical and nonclassical input states at fixed resource squeezing \cite{dega1,dega2,dega3,dega4,dega5}.

In general, the non-Gaussian character of a quantum state $\rho$ can be measured by its relative entropy distance from the Gaussian state with the same first and second moments of $\rho$ \cite{genoninong,mariannong}. It would be desirable to construct a resource theory of non-Gaussianity \cite{resourcetheories}, but to the best of our knowledge there is no clear protocol where non-Gaussianity alone plays the role of a resource which can be operationally linked to an increase in a figure of merit for some task. Certainly, this is an interesting open direction for further investigation.

Lots of theoretical effort is being put into the derivation of suitable entanglement criteria for non-Gaussian states. To detect non-Gaussian entanglement, higher order moments are often needed, and most criteria are developed around hierarchies of inequalities which check for inseparability by analysing suitable combinations of moments of arbitrary order, see e.g.~\cite{shukvog}.

On the technological side, constant improvements are being recorded for the quality of squeezing sources and detectors, which overall result in steady progresses in the experimental quality. Sometimes, breakthroughs occur, like the recent demonstration of large \cite{menicucci60} and ultra-large \cite{menicucci10k} Gaussian cluster states (of 60 and 10000 modes, respectively) in different physical domains. Gaussian cluster states are universal resources for CV one-way quantum computation \cite{menicucci} (which can be fault-tolerant even without infinite squeezing in the resource states \cite{menicuccifault}), yet the computation requires non-Gaussian measurements such as photon counting to be accomplished. This is an instance of a hybrid protocol (in this case, hybrid between Gaussian and non-Gaussian) where one needs to take the best of both worlds for superior performance.

In our opinion, hybrid routes to quantum technology are perhaps the most promising ones for the near future. Linking back to the introduction of this work, we believe that both analog and digital approaches have their merits and drawbacks, and tailored combinations can succeed to overcome selected technical issues. Recently, hybrid protocols combining CV and discrete-variable  techniques have been proposed and demonstrated. For instance, a novel (probabilistic) teleportation scheme which     `shreds' CV states and teleports them using parallel qubit channels has been proposed \cite{anderralph}, promising potentially improved performances even for the teleportation of ensembles of Gaussian states \cite{benchmark}, at finite available entanglement \cite{giannis}. Conversely, CV teleportation \cite{Braunstein98} has been recently employed to teleport deterministically the state of a single qubit \cite{furunature2013}. Hybrid models of quantum computation, where qubit degrees of
freedom for computation are combined with quantum CV buses for communication have been proposed \cite{ibridovanloock,bubu}, as well as schemes of hybrid quantum repeaters for long-distance distribution of entanglement
\cite{ripetitoreibrido}. More broadly, light-matter interfaces \cite{memorypolzik,telepolzik} involving continuous and discrete variables are essential to patching various building blocks for a future quantum internet \cite{qinternet}.

We hope this article may be useful for students approaching the field of quantum information with an interest in continuous variables. Despite many useful reviews, this important area of research has suffered from a lack of a more introductive text to get students started, saving several hours of head scratching and inefficient computations (for instance, who hasn't tried complicated integrals just to discover at the end how easy it is to apply partial traces on Gaussian states at the covariance matrix level?). We feel this article is far from filling this gap, yet it provides hopefully a step in the right direction. We anticipate that a full-length textbook on the subject is upcoming \cite{seralenuovolibro}, to which we shall warmly direct our future readers to explore all the avenues we omitted to follow in this text. For reasons of space and time, we draw this article to a conclusion.

We promised a list of possible open problems. The careful reader will have spotted two already in the previous pages. We planned to include {\it the} problem {\it per eccellenza} in CV quantum information, namely the bosonic additivity conjecture, but this problem has just been solved \cite{giovaproof1,giovaproof2,giovaproof3}, although its full implications are still to be explored. There are  a number of further questions not settled yet for Gaussian states; a small subset is, in random order: a quantitative treatment of EPR steering \cite{quantifsteering}; the design of optimal deterministic or probabilistic protocols to teleport ensembles of Gaussian states (especially squeezed ones) beating the recently set benchmarks \cite{benchmark}; a proof (or disproof) of strong monogamy \cite{strongmono} of Gaussian entanglement for arbitrary nonsymmetric $N$-mode Gaussian states; etc. There are certainly many more open problems related to specific protocols, e.g. quantum key distribution and communication in general \cite{pirandolareview}, as well as applications and developments to other research areas, such as relativistic quantum information  \cite{nicoiv,cqg,antthesis,nicothesis,antteleprl}. We prefer to leave our curious readers to find their own ones to try and tackle.

\section*{Acknowledgments}
We are pleased to acknowledge a number of people with whom we worked in recent years on topics related to this article. In alphabetical order, our thanks go in particular to N. Friis, I. Fuentes, D. Girolami, F. Illuminati, L. Mi\v{s}ta Jr., and  A. Serafini. GA is grateful to S. Pascazio for the kind invitation to contribute this article, his endured tolerance with deadlines, and his careful reading of the manuscript. Apologies are issued {\it a priori} to whom it may concern for all the missing references. Financial support from the University of Nottingham [EPSRC Doctoral Prize], the Foundational Questions Institute [Grant No. FQXi-RFP3-1317] and the Brazilian  CAPES [Pesquisador Visitante Especial-Grant No. 108/2012] is gratefully acknowledged.


\begin{thebibliography}{100}

\bibitem{BNielsChuang}
M.~A. Nielsen and I.~L. Chuang, \emph{Quantum Computation and Quantum
  Information}, Cambridge University Press, Cambridge, 2000.

\bibitem{qinternet}
H.~J. Kimble, Nature \textbf{453} (2008), no.~7198, 1023.

\bibitem{BB84}
C.~H. Bennett and G.~Brassard, \emph{Quantum cryptography: Public key
  distribution and coin tossing}, Proceedings of the IEEE International
  Conference on Computers, Systems and Signal Processing, Bangalore (1984),
  p.~175.

\bibitem{Ekert91}
A.~K. Ekert, Phys. Rev. Lett. \textbf{67} (1991), 661--663.

\bibitem{Telep}
C.~H. Bennett, G.~Brassard, C.~Cr\'{e}peau, R.~Jozsa, A.~Peres, and W.~K.
  Wootters, Phys. Rev. Lett. \textbf{70} (1993), 1895.

\bibitem{idquantique}
{{ID Quantique}}, \underline{\texttt{http://www.idquantique.com/}}.

\bibitem{dwave}
S.~Boixo, T.~Albash, F.~M. Spedalieri, N.~Chancellor, and D.~A. Lidar, Nat.
  Commun. \textbf{4} (2013), 2067.

\bibitem{sammyrev}
B.~L. Schumaker, Phys. Rep. \textbf{135} (1986), 317.

\bibitem{eisplenio}
J.~Eisert and M.~B. Plenio, Int. J. Quant. Inf. \textbf{1} (2003), 479.

\bibitem{BBraunstein}
S.~L. Braunstein and A.~K. Pati eds., \emph{Quantum Information with Continuous
  Variables}, Kluwer Academic, Dordrecht, 2003.

\bibitem{parisbook}
A.~Ferraro, S.~Olivares, and M.~G.~A. Paris, \emph{Gaussian states in quantum
  in continuous variable quantum information}, Bibliopolis, Napoli, 2005.

\bibitem{brareview}
S.~L. Braunstein and P.~{van Loock}, Rev. Mod. Phys. \textbf{77} (2005), 513.

\bibitem{book}
N.~Cerf, G.~Leuchs, and E.~S. Polzik eds., \emph{Quantum Information with
  Continuous Variables of Atoms and Light}, Imperial College Press, London,
  2007.

\bibitem{ourreview}
G.~Adesso and F.~Illuminati, J. Phys. A: Math. Theor. \textbf{40} (2007), 7821.

\bibitem{ciracreview}
N.~Schuch, J.~I. Cirac, and M.~M. Wolf, Commun. Math. Phys. \textbf{267}
  (2006), 65.

\bibitem{pirandolareview}
C. Weedbrook, S. Pirandola, R. Garcia-Patron, N. J. Cerf, T. C. Ralph, J. H. Shapiro, and S. Lloyd,
Rev. Mod. Phys. \textbf{84} (2012), 621.

\bibitem{extra}
M.~M. Wolf, G.~Giedke, and J.~I. Cirac, Phys. Rev. Lett. \textbf{96} (2006),
  080502.

\bibitem{Arvind95}
Arvind, B.~Dutta, N.~Mukunda, and R.~Simon, Pramana \textbf{45} (1995), 471.

\bibitem{BWallsMilburn}
D.~F. Walls and G.~J. Milburn, \emph{Quantum Optics}, Springer--Verlag, Berlin,
  1995.

\bibitem{barnett}
A.~M. Barnett and P.~M. Radmore, \emph{Methods in Theoretical Quantum Optics},
  Clarendon Press, Oxford, 1997.

\bibitem{Husimi40}
K.~Husimi, Proc. Phys. Math. Soc. Jpn \textbf{23} (1940), 264.

\bibitem{Wigner32}
E.~P. Wigner, Phys. Rev. \textbf{40} (1932), 749.

\bibitem{Glauber63}
R.~J. Glauber, Phys. Rev. \textbf{131} (1963), 2766.

\bibitem{Sudarshan63}
E.~C.~G. Sudarshan, Phys. Rev. Lett. \textbf{10} (1963), 277.

\bibitem{bellrev}
N.~Brunner, D.~Cavalcanti, S.~Pironio, V.~Scarani, and S.~Wehner (2013), Rev.
  Mod. Phys. to appear, arXiv:1303.2849.

\bibitem{entanglement}
R.~Horodecki, P.~Horodecki, M.~Horodecki, and K.~Horodecki, Rev. Mod. Phys.
  \textbf{81} (2009), 865.

\bibitem{modireview}
K.~Modi, A.~Brodutch, H.~Cable, T.~Paterek, and V.~Vedral, Rev. Mod. Phys.
  \textbf{84} (2012), 1655.

\bibitem{ParisFerraro}
A.~Ferraro and M.~G.~A. Paris, Phys. Rev. Lett. \textbf{108} (2012), 260403.

\bibitem{cahillglau1}
K.~E. Cahill and R.~J. Glauber, Phys. Rev. \textbf{177} (1969), 1857.

\bibitem{cahillglau2}
K.~E. Cahill and R.~J. Glauber, Phys. Rev. \textbf{177} (1969), 1882.

\bibitem{Simon00}
R.~Simon, Phys. Rev. Lett. \textbf{84} (2000), 2726.

\bibitem{sculzub}
M.~O. Scully and M.~S. Zubairy, \emph{Quantum Optics}, Cambridge University
  Press, Cambridge, 1997.

\bibitem{francamentemeneinfischio}
J.~Laurat, G.~Keller, J.~A. {Oliveira-Huguenin}, C.~Fabre, T.~Coudreau,
  A.~Serafini, G.~Adesso, and F.~Illuminati, J. Opt. B: Quantum Semiclass. Opt.
  \textbf{7} (2005), S577.

\bibitem{simon87}
R.~Simon, E.~C.~G. Sudarshan, and N.~Mukunda, Phys. Rev. A \textbf{36} (1987),
  3868.

\bibitem{simon94}
R.~Simon, N.~Mukunda, and B.~Dutta, Phys. Rev. A \textbf{49} (1994), 1567.

\bibitem{robertson30}
H.~P. Robertson, Phys. Rev. \textbf{34} (1929), 163.

\bibitem{schrodinger30}
E.~Schr{\"o}dinger, Berg. Kgl. Akad Wiss. (1930), 296.

\bibitem{serafozziprl}
A.~Serafini, Phys. Rev. Lett. \textbf{96} (2006), 110402.

\bibitem{Hudson}
R.~Hudson, Rep. Math. Phys. \textbf{6} (1974), no.~2, 249 -- 252.

\bibitem{Mandilara}
A.~Mandilara, E.~Karpov, and N.~J. Cerf, Phys. Rev. A \textbf{79} (2009),
  062302.

\bibitem{Filip111}
R.~Filip and L.~Mi\ifmmode~\check{s}\else \v{s}\fi{}ta, Phys. Rev. Lett.
  \textbf{106} (2011), 200401.

\bibitem{Filip112}
M.~Je\v{z}ek, I.~Straka, M.~Mi\v{c}uda, M.~Du\v{s}ek, J.~Fiur\'a\v{s}ek, and
  R.~Filip, Phys. Rev. Lett. \textbf{107} (2011), 213602.

\bibitem{Genoni2013}
M.~G. Genoni, M.~L. Palma, T.~Tufarelli, S.~Olivares, M.~S. Kim, and M.~G.~A.
  Paris, Phys. Rev. A \textbf{87} (2013), 062104.

\bibitem{berndt2001}
R.~Berndt, \emph{An Introduction to Symplectic Geometry}, American Mathematical
  Soc., 2001.

\bibitem{gosson2006}
M.~A. {de Gosson}, \emph{Symplectic Geometry and Quantum Mechanics},
  Birkh{\"a}user Basel, 2006.

\bibitem{williamson}
J.~Williamson, Am. J. Math. \textbf{58} (1936), 141.

\bibitem{SeralePHD}
A.~Serafini, Ph.D. thesis\ (Universit\`a degli Studi di Salerno, 2004), \\
  \texttt{www.tampa.phys.ucl.ac.uk/quinfo/people/alessiothesis.pdf}.

\bibitem{generic}
G.~Adesso, Phys. Rev. Lett. \textbf{97} (2006), 130502.

\bibitem{antthesis}
A.~R. Lee, Ph.D. thesis\ (University of Nottingham, 2013), arXiv:1309.4419.

\bibitem{hall2004}
B.~Hall, \emph{Lie Groups, Lie Algebras, and Representations: An Elementary
  Introduction}, Springer, 2003.

\bibitem{luis1995}
A.~Luis and L.~L. Sanchez-Soto, Quantum and Semiclassical Optics: Journal of
  the European Optical Society Part B \textbf{7} (1995), no.~2, 153+.

\bibitem{serafozzinazi}
A.~Serafini, J.~Eisert, and M.~M. Wolf, Phys. Rev. A \textbf{71} (2005),
  012320.

\bibitem{braunsqueezirreducibile}
S.~L. Braunstein, Phys. Rev. A \textbf{71} (2005), 055801.

\bibitem{serafozzijob05}
A.~Serafini, M.~G.~A. Paris, F.~Illuminati, and S.~{De Siena}, J. Opt. B:
  Quantum Semiclass. Opt. \textbf{71} (2005), R19.

\bibitem{wolfeis}
M.~M. Wolf and J.~Eisert, New J. Phys. \textbf{7} (2005), 93.

\bibitem{holevonew}
F.~Caruso, J.~Eisert, V.~Giovannetti, and A.~Holevo, New J. Phys. \textbf{10}
  (2008), 083030.

\bibitem{mistagauss}
L.~Mi\v{s}ta, R.~Tatham, D.~Girolami, N.~Korolkova, and G.~Adesso, Phys. Rev. A
  \textbf{83} (2011), 042325.

\bibitem{fiurasek07}
J.~Fiur\'a\v{s}ek and L.~Mi\v{s}ta, Phys. Rev. A \textbf{75} (2007), 060302.

\bibitem{nogo1}
J.~Eisert, S.~Scheel, and M.~B. Plenio, Phys. Rev. Lett. \textbf{89} (2002),
  137903.

\bibitem{nogo2}
J.~Fiur\'a\u{s}ek, Phys. Rev. Lett. \textbf{89} (2002), 137904.

\bibitem{nogo3}
G.~Giedke and J.~I. Cirac, Phys. Rev. A \textbf{66} (2002), 032316.

\bibitem{EPR35}
A.~Einstein, B.~Podolsky, and N.~Rosen, Phys. Rev. \textbf{47} (1935), 777.

\bibitem{stable10db}
T.~Eberle, V.~H\"andchen, and R.~Schnabel, Opt. Expr. \textbf{21} (2013),
  11546.

\bibitem{vaidman}
L.~Vaidman, Phys. Rev. A \textbf{49} (1994), 1473.

\bibitem{Braunstein98}
S.~L. Braunstein and H.~J. Kimble, Phys. Rev. Lett. \textbf{80} (1998), 869.

\bibitem{Furusawa98}
A.~Furusawa, J.~L. S{\o}rensen, S.~L. Braunstein, C.~A. Fuchs, H.~J. Kimble,
  and E.~S. Polzik, Science \textbf{282} (1998), 706.

\bibitem{vanlokfortshit}
P.~{van Loock}, Fortschr. Phys. \textbf{50} (2002), 12 1177.

\bibitem{zykbook}
I.~Bengtsson and K.~Zyczkowski, \emph{Geometry of Quantum States: An
  Introduction to Quantum Entanglement}, Cambridge University Press, Cambridge,
  2006.

\bibitem{holevobound}
A.~S. Holevo, Probl. Inf. Transm. \textbf{9} (1973), 177.

\bibitem{merging}
M.~Horodecki, J.~Oppenheim, and A.~Winter, Nature \textbf{436} (2005), 673.

\bibitem{mother}
I.~Devetak, A.~W. Harrow, and A.~Winter, Phys. Rev. Lett. \textbf{93} (2004),
  230503.

\bibitem{proofsss}
E.~H. Lieb and M.~B. Ruskai, J. Math. Phys. \textbf{14} (1973), 1938.

\bibitem{hammer}
M.~A. Nielsen and D.~Petz, Quant. Inf. Comput. \textbf{5} (2005), 507.

\bibitem{Wehrl78}
A.~Wehrl, Rev. Mod. Phys. \textbf{50} (1978), 221.

\bibitem{Renyi1970Book}
A.~R{\'e}nyi, \emph{Probability theory}, Elsevier, 1970.

\bibitem{baez}
J.~C. Baez, arXiv:1102.2098 (2011).

\bibitem{renyicapacity}
M.~Mosonyi and F.~Hiai, IEEE Trans. Inf. Th. \textbf{57} (2011), 2474.

\bibitem{workvalue}
O.~C.~O. Dahlsten, R.~Renner, E.~Rieper, and V.~Vedral, New J. Phys.
  \textbf{13} (2011), 053015.

\bibitem{renyispectrum}
F.~Franchini, A.~R. Its, and V.~E. Korepin, J. Phys. A: Math. Theor.
  \textbf{41} (2008), 025302.

\bibitem{extremal}
G.~Adesso, A.~Serafini, and F.~Illuminati, Phys. Rev. A \textbf{70} (2004),
  022318.

\bibitem{renyi}
G.~Adesso, D.~Girolami, and A.~Serafini, Phys. Rev. Lett. \textbf{109} (2002),
  190502.

\bibitem{Shannon48}
C.~E. Shannon, Bell Syst. Tech. J. \textbf{27} (1948), 379.

\bibitem{hornjohnson}
R.~A. Horn and C.~R. Johnson, \emph{Matrix Analysis}, Cambridge University
  Press, Cambridge, 1990.

\bibitem{vidwer02}
G.~Vidal and R.~F. Werner, Phys. Rev. A \textbf{65} (2002), 032314.

\bibitem{peres96}
A.~Peres, Phys. Rev. Lett. \textbf{77} (1996), 1413.

\bibitem{horodecki96}
M.~Horodecki, P.~Horodecki, and R.~Horodecki, Phys. Lett. A \textbf{223}
  (1996), 1.

\bibitem{Bennett96pra}
C.~H. Bennett, D.~P. DiVincenzo, J.~A. Smolin, and W.~K. Wootters, Phys. Rev. A
  \textbf{54} (1996), 3824.

\bibitem{geof}
M.~M. Wolf, G.~Giedke, O.~Kr{\"u}ger, R.~F. Werner, and J.~I. Cirac, Phys. Rev.
  A \textbf{69} (2004), 052320.

\bibitem{ordering}
G.~Adesso and F.~Illuminati, Phys. Rev. A \textbf{72} (2005), 032334.

\bibitem{Holevo01}
A.~S. Holevo and R.~F. Werner, Phys. Rev. A \textbf{63} (2001), 032312.

\bibitem{giedke03}
G.~Giedke, M.~M. Wolf, O.~Kr\"uger, R.~F. Werner, and J.~I. Cirac, Phys. Rev.
  Lett. \textbf{91} (2003), 107901.

\bibitem{giovaproof1}
V.~Giovannetti, A.~S. Holevo, and R.~Garcia-Patron, arXiv:1312.2251 (2013).

\bibitem{giovaproof2}
A.~Mari, V.~Giovannetti, and A.~S. Holevo, arXiv:1312.3545 (2013).

\bibitem{giovaproof3}
V.~Giovannetti, R.~Garcia-Patron, N.~J. Cerf, and A.~S. Holevo, arXiv:1312.6225
  (2013).

\bibitem{zurek}
H.~Ollivier and W.~H. Zurek, Phys. Rev. Lett. \textbf{88} (2001), 017901.

\bibitem{vedral}
L.~Henderson and V.~Vedral, J. Phys. A: Math. Gen. \textbf{34} (2001), 6899.

\bibitem{adessodatta}
G.~Adesso and A.~Datta, Phys. Rev. Lett. \textbf{105} (2010), 030501.

\bibitem{giordaparis}
P.~Giorda and M.~G.~A. Paris, Phys. Rev. Lett. \textbf{105} (2010), 020503.

\bibitem{ralphdiscord}
S.~Rahimi-Keshari, C.~M. Caves, and T.~C. Ralph, Phys. Rev. A \textbf{87}
  (2013), 012119.

\bibitem{allegramente}
P.~Giorda, M.~Allegra, and M.~G.~A. Paris, Phys. Rev. A \textbf{86} (2012),
  052328.

\bibitem{pirlaimplications}
S.~Pirandola, N.~J. Cerf, S.~L. Braunstein, and S.~Lloyd, arXiv:1309.2215
  (2013).

\bibitem{Duan00}
L.-M. Duan, G.~Giedke, J.~I. Cirac, and P.~Zoller, Phys. Rev. Lett. \textbf{84}
  (2000), 2722.

\bibitem{Serafini2004JPB}
A.~Serafini, F.~Illuminati, and S.~D. Siena, Journal of Physics B: Atomic,
  Molecular and Optical Physics \textbf{37} (2004), no.~2, L21.

\bibitem{ckw}
V.~Coffman, J.~Kundu, and W.~K. Wootters, Phys. Rev. A \textbf{61} (2000),
  052306.

\bibitem{pisa}
G.~Adesso and F.~Illuminati, Int. J. Quant. Inf. \textbf{4} (2006), 383.

\bibitem{barrylagallina}
J.~S. Kim and B.~C. Sanders, J. Phys. A: Math. Theor. \textbf{43} (2010),
  445305.

\bibitem{osborne}
T.~J. Osborne and F.~Verstraete, Phys. Rev. Lett. \textbf{96} (2006), 220503.

\bibitem{hiroshimaa}
T.~Hiroshima, G.~Adesso, and F.~Illuminati, Phys. Rev. Lett. \textbf{98}
  (2007), 050503.

\bibitem{contangle}
G.~Adesso and F.~Illuminati, New J. Phys. \textbf{8} (2006), 15.

\bibitem{3mpra}
G.~Adesso, A.~Serafini, and F.~Illuminati, Phys. Rev. A \textbf{73} (2006),
  032345.

\bibitem{samy}
G.~Adesso and S.~Piano, Phys. Rev. Lett. \textbf{112} (2014), 010401.

\bibitem{strongmono}
G.~Adesso and F.~Illuminati, Phys. Rev. Lett. \textbf{99} (2007), 150501.

\bibitem{cerfbit}
L.~Magnin, F.~Magniez, A.~Leverrier, and N.~J. Cerf, Phys. Rev. A \textbf{81}
  (2010), 010302.

\bibitem{nogoerr}
J.~Niset, J.~Fiur\'a\v{s}ek, and N.~J. Cerf, Phys. Rev. Lett. \textbf{102}
  (2009), 120501.

\bibitem{mypraest}
G.~Adesso, F.~Dell'Anno, S.~De~Siena, F.~Illuminati, and L.~A.~M. Souza, Phys.
  Rev. A \textbf{79} (2009), 040305.

\bibitem{nongaussexp}
A.~Ourjoumtsev, A.~Dantan, R.~Tualle-Brouri, and P.~Grangier, Phys. Rev. Lett.
  \textbf{98} (2007), 030502.

\bibitem{furunphot2010}
H.~Takahashi, J.~S. Neergaard-Nielsen, M.~Takeuchi, M.~Takeoka, K.~Hayasaka,
  and A.~Furusawa, Nat. Photon. \textbf{4} (2010), 178.

\bibitem{dega1}
T.~Opatrny, G.~Kurizki, and D.-G. Welsch, Phys. Rev. A \textbf{61} (2000),
  032302.

\bibitem{dega2}
P.~T. Cochrane, T.~C. Ralph, and G.~J. Milburn, Phys. Rev. A \textbf{65}
  (2002), 062306.

\bibitem{dega3}
S.~Olivares, M.~G.~A. Paris, and R.~Bonifacio, Phys. Rev. A \textbf{67} (2003),
  032314.

\bibitem{dega4}
A.~Kitagawa, M.~Takeoka, M.~Sasaki, and A.~Chefles, Phys. Rev. A \textbf{73}
  (2006), 042310.

\bibitem{dega5}
F.~Dell'Anno, S.~D. Siena, L.~A. Farias, and F.~Illuminati, Phys. Rev. A
  \textbf{76} (2007), 022301.

\bibitem{genoninong}
M.~G. Genoni, M.~G.~A. Paris, and K.~Banaszek, Phys. Rev. A \textbf{78} (2008),
  060303.

\bibitem{mariannong}
P.~Marian and T.~A. Marian, Phys. Rev. A \textbf{88} (2013), 012322.

\bibitem{resourcetheories}
M.~Horodecki and J.~Oppenheim, Int. J. Mod. Phys. B \textbf{27} (2013),
  1345019.

\bibitem{shukvog}
E.~Shchukin and W.~Vogel, Phys. Rev. Lett. \textbf{95} (2005), 230502.

\bibitem{menicucci60}
M.~Chen, N.~C. Menicucci, and O.~Pfister, arXiv:1311.2957 (2013).

\bibitem{menicucci10k}
S.~Yokoyama, R.~Ukai, S.~C. Armstrong, C.~Sornphiphatphong, T.~Kaji, S.~Suzuki,
  J.-I. Yoshikawa, H.~Yonezawa, N.~C. Menicucci, and A.~Furusawa, Nat. Photon.
  \textbf{7} (2013), 982.

\bibitem{menicucci}
N.~C. Menicucci, P.~{van Loock}, M.~Gu, C.~Weedbrook, T.~C. Ralph, and M.~A.
  Nielsen, Phys. Rev. Lett. \textbf{97} (2005), 110501.

\bibitem{menicuccifault}
N.~C. Menicucci, arXiv:1310.7596 (2013).

\bibitem{anderralph}
U.~L. Andersen and T.~C. Ralph, Phys. Rev. Lett. \textbf{111} (2013), 050504.

\bibitem{benchmark}
G.~Chiribella and G.~Adesso, Phys. Rev. Lett. \textbf{112} (2014), 010501.

\bibitem{giannis}
I.~Kogias, S.~Ragy, and G.~Adesso (2014), in preparation.

\bibitem{furunature2013}
S.~Takeda, T.~Mizuta, M.~Fuwa, P.~{van Loock}, and A.~Furusawa, Nature
  \textbf{500} (2013), 315.

\bibitem{ibridovanloock}
T.~P. Spiller, K.~Nemoto, S.~L. Braunstein, W.~J. Munro, P.~{van Loock}, and
  G.~J. Milburn, New J. Phys. \textbf{8} (2006), 30.

\bibitem{bubu}
P.~{van Loock}, W.~J. Munro, K.~Nemoto, T.~P. Spiller, T.~D. Ladd, S.~L.
  Braunstein, and G.~J. Milburn, Phys. Rev. A \textbf{78} (2008), 022303.

\bibitem{ripetitoreibrido}
P.~van Loock, T.~D. Ladd, K.~Sanaka, F.~Yamaguchi, K.~Nemoto, W.~J. Munro, and
  Y.~Yamamoto, Phys. Rev. Lett. \textbf{96} (2006), 240501.

\bibitem{memorypolzik}
B.~Julsgaard, J.~Sherson, J.~I. Cirac, J.~Fiur\'a\u{s}ek, and E.~S. Polzik,
  Nature \textbf{432} (2004), 482.

\bibitem{telepolzik}
J.~F. Sherson, H.~Krauter, R.~K. Olsson, B.~Julsgaard, K.~Hammerer, J.~I.
  Cirac, and E.~S. Polzik, Nature \textbf{443} (2006), 557.

\bibitem{seralenuovolibro}
A.~Serafini and V.~Giovannetti (2014), textbook in preparation.

\bibitem{quantifsteering}
P.~Skrzypczyk, M.~Navascues, and D.~Cavalcanti, arXiv:1311.4590 (2013).

\bibitem{nicoiv}
N.~Friis and I.~Fuentes, J. Mod. Opt. \textbf{60} (2013), 22.

\bibitem{cqg}
G.~Adesso, S.~Ragy, and D.~Girolami, Class. Quantum Grav. \textbf{29} (2012),
  224002.

\bibitem{nicothesis}
N.~Friis, Ph.D. thesis\ (University of Nottingham, 2013), arXiv:1311.3536.

\bibitem{antteleprl}
N.~Friis, A.~R. Lee, K.~Truong, C.~Sab\'in, E.~Solano, G.~Johansson, and
  I.~Fuentes, Phys. Rev. Lett. \textbf{110} (2013), 113602.

\end{thebibliography}

\providecommand{\bysame}{\leavevmode\hbox to3em{\hrulefill}\thinspace}
\providecommand{\MR}{\relax\ifhmode\unskip\space\fi MR }
\providecommand{\MRhref}[2]{%
  \href{http://www.ams.org/mathscinet-getitem?mr=#1}{#2}
}
\providecommand{\href}[2]{#2}

\end{document}